\documentclass[final,1p,twocolumn]{elsarticle}

\usepackage[numbers]{natbib}
\usepackage{amssymb}
\usepackage{graphicx}
\usepackage{dcolumn}
\usepackage{bm}
\usepackage{hyperref}
\usepackage{subfigure}
\usepackage{xcolor}

\journal{Applied Mathematics and Computation}

\begin{document}

\begin{frontmatter}

\title{Dynamical Chaos in a Simple Model of a Knuckleball}

\author{Nicholas J. Nelson}
\ead{njnelson@csuchico.edu}

\author{Eric Strauss}
\ead{estrauss1@mail.csuchico.edu}
\address{Department of Physics \\
 California State University, Chico \\
 400 W. 1st St. \\
 Chico, CA 95929}


\begin{abstract}
The knuckleball is perhaps the most enigmatic pitch in baseball. Relying on the presence of raised seams on the surface of the ball to create asymmetric flow, a knuckleball's trajectory has proven very challenging to predict compared to other baseball pitches, such as fastballs or curveballs. Previous experimental tracking of large numbers of knuckleballs has shown that they can move in essentially any direction relative to what would be expected from a drag-only trajectory. This has led to speculation that knuckleballs exhibit chaotic motion. Here we develop a relatively simple model of a knuckleball that includes quadratic drag and lift from asymmetric flow which is taken from experimental measurements of slowly rotating baseballs. Our models can indeed exhibit dynamical chaos as long In contrast, models that omit torques on the ball in flight do not show chaotic behavior. Uncertainties in the phase space position of the knuckleball are shown to grow by factors as large as $10^6$ over the flight of the ball from the pitcher to home plate. We quantify the impact of our model parameters on the chaos realized in our models, specifically showing that maximum Lyapunov exponent is roughly proportional to the square root of the effective lever arm of the torque, and also roughly proportional to the initial velocity of the pitch.  We demonstrate the existence of bifurcations that can produce changes in the location of the ball when it reaches the plate of as much as 1.2 m for specific initial conditions similar to those used by professional knuckleball pitchers. As we introduce additional complexity in the form of more faithful representations of the empirical asymmetry force measurements, we find that a larger fraction of the possible initial conditions result in dynamical chaos. The chaotic behavior seen in our simplified model combined with the additional complex and nonlinear effects likely present in real knuckleballs provide strong evidence that knuckleballs are in fact chaotic, and that the chaotic uncertainty is likely a significant factor in the unpredictability of the pitch.
\end{abstract}

\end{frontmatter}

\section{\label{sec:intro} ``Throwing a butterfly with hiccups''}

The knuckleball is one of the most specialized pitches in baseball. Most pitchers try to confuse batters by throwing a variety of pitches such as fastballs, sliders, and curveballs, each with a unique trajectory achieved by changing the speed and rotation of the ball. Knuckleballers, however, achieve confusion by repeatedly throwing the same pitch. The knuckleball, with its extremely slow rotation in flight, produces a trajectory that is difficult for pitchers, catchers, and hitters alike to predict. Major League Baseball Hall-of-Famer Willie Stargel once stated that ``throwing a knuckleball for a strike is like throwing a butterfly with hiccups across the street into your neighbor's mailbox'' \cite{Clark2012}. 

Knuckleballs are thrown in such a way as to minimize the rotation of the ball once it has left the pitcher's hand. In the words of long-time knuckleball pitcher Tim Wakefield, ``If it spins at all it basically doesn't work.'' \cite{Wakefield2011} Baseballs have a pattern of raised seams, which break the symmetry of the flow of air around the ball and result in aerodynamic forces the depend upon the orientation of the ball relative to its direction of motion. As the ball slowly spins these forces change in both magnitude and direction, producing relatively large changes in the ball's vertical and horizontal velocities as it travels. This makes knuckleballs very difficult to hit. Similar behaviors are seen in other sports, such as volleyball and soccer, though in those cases the asymmetric flow is likely related to hydrodynamic instabilities in the flow around the ball.

Knuckleballs have a reputation in baseball as being sensitive to how they are thrown down to levels beyond the pitchers ability to control. ``You can throw two knuckleballs with the identical release, the identical motion, in the identical place, and one might go one way and the second might go another way'' is how professional knuckleball pitcher R.A. Dickey describes them \cite{Dickey2012}. Due to the complex motions observed in individual pitches and perhaps to the fact that everyone involved with the pitch is unable to predict its motion, some have speculated that kunckleballs may exhibit dynamical chaos. Dynamical chaos is seen in many physical systems \cite[e.g.,][]{Lorenz1963, Perc2005, Strogatz2015, Brown2018, Nepomuceno2020} and is defined by a strong sensitivity to initial conditions. More precisely, dynamical chaos is seen when a system shows exponential divergence in time for two sets of nearly identical initial conditions. Other hallmarks include bifurcations where the changing of a parameter by an infinitesimal amount results in qualitatively different behavior, and intermittency, where chaotic regions in parameter space are interspersed with regions that display non-chaotic behaviors. In this paper we will show that indeed knuckleballs do show clear evidence of chaotic behaviors, even for our simplified models, as long as those models admit variation in the rotation rate of the ball.

In this paper we will develop a set of simple models for the trajectory of a knuckleball as it travels from the pitcher's hand to home plate. We will then explore solutions of these models and show that they yield dynamical chaos as long as torques on the ball are permitted. Chaotic behavior is realized when only one axis of rotation is considered, as well as the more realistic case when two axes of rotation are allowed. The time-scale for the growth of uncertainty will be shown to depend on both the effective lever arm of the torque, and the speed of the pitch.  We will further show that simply by altering the initial angle and angular velocity of the ball by very small amounts, a pitch with identical initial position and velocity can arrive at the plate at a wide range of locations.

\section{\label{sec:model} Developing a Knuckleball Model}

In this work we follow the common convention used by Major League Baseball's StatCast system, which sets a coordinate system where home plate in located at the origin \cite{Baseball2019}. The $+\hat{y}$ direction points from home plate to the pitcher's mound, the $+\hat{z}$ direction is upward, and the $+\hat{x}$ direction points to the pitcher's left (batter's right). The strike zone is a rectangular portion of the $x$-$z$ plane that is bounded on the horizontal edges by the sides of home plate located at $x = \pm 0.216$ m. The vertical limits of the strike zone are defined based on the positions of the shoulders, waist, and knees of the batter as they prepare to swing. Here we choose to adopt a strike zone that extends from $z = 0.53$ m to $z = 1.19$ m, based on long-term averages of called strikes in Major League Baseball games \cite{Roegele2014}.

The trajectory of an object moving through a fluid (such as air) is a classic problem encountered by most physics students in their first physics class. To determine the trajectory taken, one must identify and quantify all of the relevant forces on the object and then integrate in time. The aerodynamic forces on a baseball have been studied extensively both in laboratory settings and using tracking systems deployed in Major League Baseball games \cite{Watts1987, Nathan2008, Nathan2012, Kagan2014, EscaleraSantos2019}. In addition to gravity, most baseballs in flight experience significant drag forces as well as Magnus or lift forces due to the rotation of the ball. Most thrown or batted balls show rotation rates on the order of 100 rad/s \cite{EscaleraSantos2019}. Knuckleballs, however, rotate less than about 20 rad/s. In this regime the Magnus force is extremely weak and can be ignored, however an additional force arises due to the changes in flow around the ball caused by the relative locations of the raised seams. We will call this the asymmetry force. Thus the acceleration of a knuckleball can simply be written as
\begin{equation}
    \vec{a} = \frac{\vec{F}_G + \vec{F}_D + \vec{F}_A}{m} \; ,
    \label{eq:2nd}
\end{equation}
where $m$ is the mass of the baseball. The gravitational force is the simplest of the three with $\vec{F}_G = - m g \hat{z}$ and $g$ the standard 9.8 m/s$^2$. The drag force is only slightly more complex with a functional form given by
\begin{equation}
    \vec{F}_D = - \frac{1}{2} \rho \pi R^2 C_D v \vec{v} \; ,
    \label{eq:drag}
\end{equation}
where $\rho$ is the density of the air, $R$ is the radius of the baseball, $v$ is the magnitude of the velocity, $\vec{v}$ is the velocity vector, and $C_D$ is a dimensionless coefficient that depends on the details of the flow around the ball. This drag coefficient has been extensively studied for baseballs in both laboratory settings and using data from professional baseball games \cite{Adair2002, Kagan2014, EscaleraSantos2019}. While there is some weak dependence on velocity for the range of speeds common in pitches, we have chosen to use a fixed value of $C_D = 0.346$ \cite{Kagan2014}. Our results are qualitatively unchanged using values of $C_D$ between 0.32 and 0.37.

\begin{figure}[t]
\centering
\includegraphics[width=0.5\linewidth]{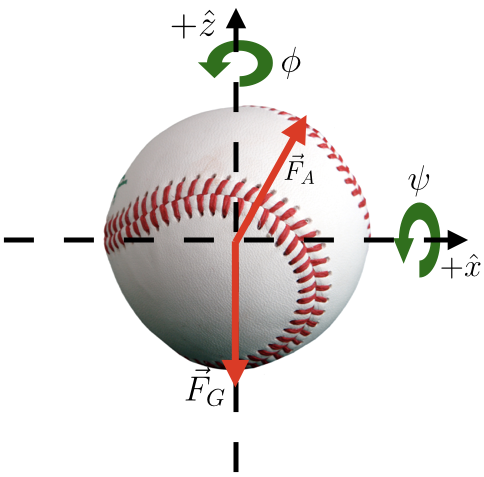}
\caption{\label{fig:forces} Schematic diagram of the forces acting on a knuckleball moving directly out of the page. The drag force (not shown) would be directed purely into the page. Also noted are the angles $\psi$ and $\phi$ which indicate rotation about the $\hat{x}$ and $\hat{z}$ axes, respectively.}
\end{figure}

The force due the asymmetric flow of air around the ball caused by the location of the seams is perhaps the most challenging. Functionally it takes the form
\begin{equation}
    \vec{F}_A = \frac{1}{2} \rho \pi R^2 C_A v^2 \; \hat{\theta} \times \hat{v} \; ,
    \label{eq:F_A}
\end{equation}
where $C_A$ is a non-dimensional coefficient and $\hat{\theta}$ is a unit vector fixed on a reference point on the ball. The directional dependence can be simplified using the approximation that the velocity vector is largely in the $\hat{y}$ direction, so $\hat{\theta} \times \hat{v}$ must have components in only the $\hat{x}$ and $\hat{z}$ directions. The functional form of Eqn.~\ref{eq:F_A} has been verified in a variety of settings \cite{Watts1975, Morrissey2009, Asai2011, Nathan2012, Higuchi2012, Borg2014, Texier2016, Aguirre-Lopez2017}, however measurement of $C_A$ has proven quite difficult. $C_A$ depends on the orientation of the ball $\vec{\theta}$, the angular velocity vector $\vec{\omega}$, and velocity vector $\vec{v}$. The best available data comes from Morrissey \cite{Morrissey2009}, who conducted detailed measurements using wind tunnel experiments for non-rotating and slowly rotating balls at a fixed velocity over the full rotation of the ball for two orientations -- the two-seam orientation and the four-seam orientation. The knuckleball is commonly thrown so that the the two-seam orientation rotates around the x-axis and the four-seam orientation rotates about the z-axis \citep{Clark2012}.

\begin{figure*}[t]
\includegraphics[width=0.49\linewidth]{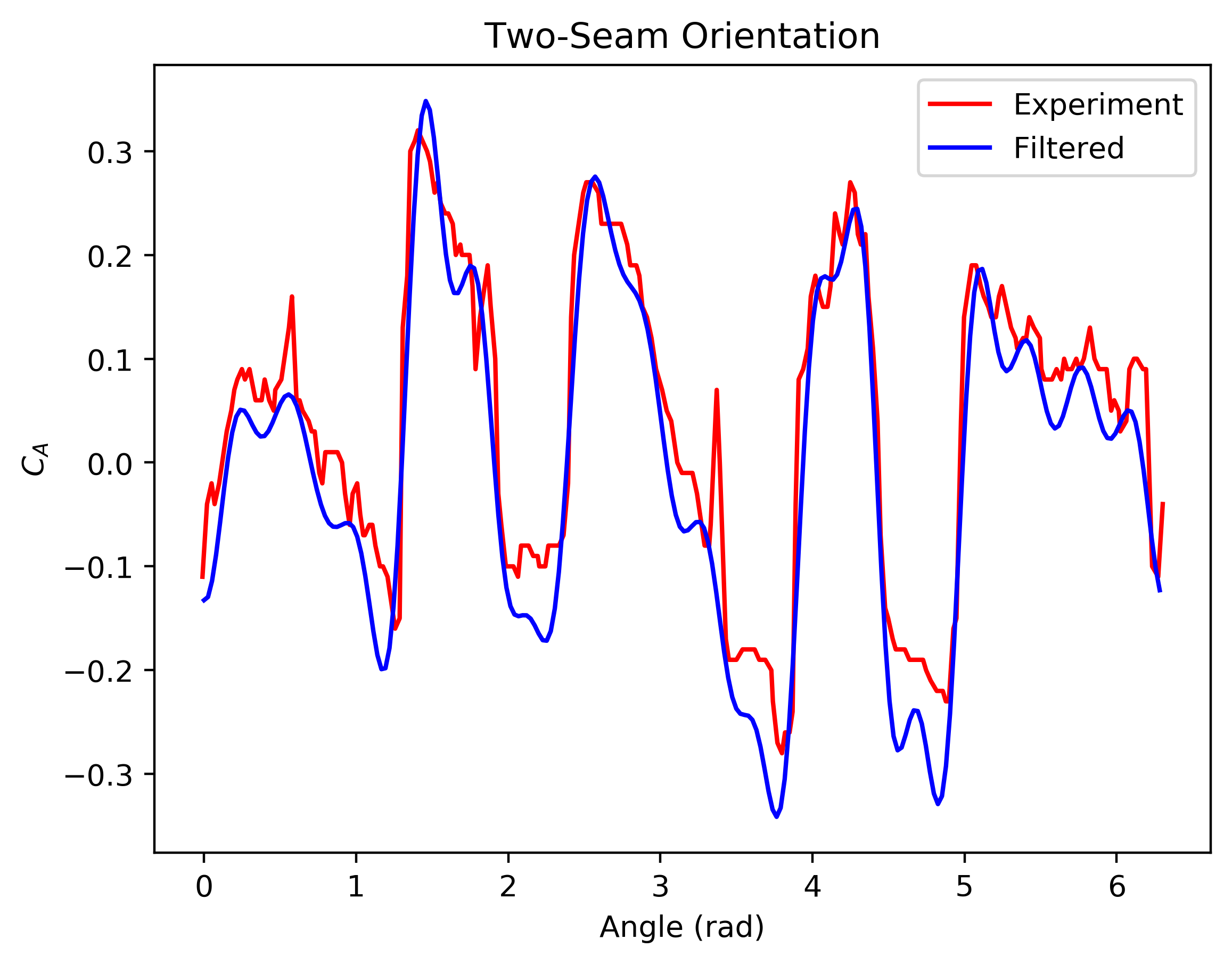}
\includegraphics[width=0.49\linewidth]{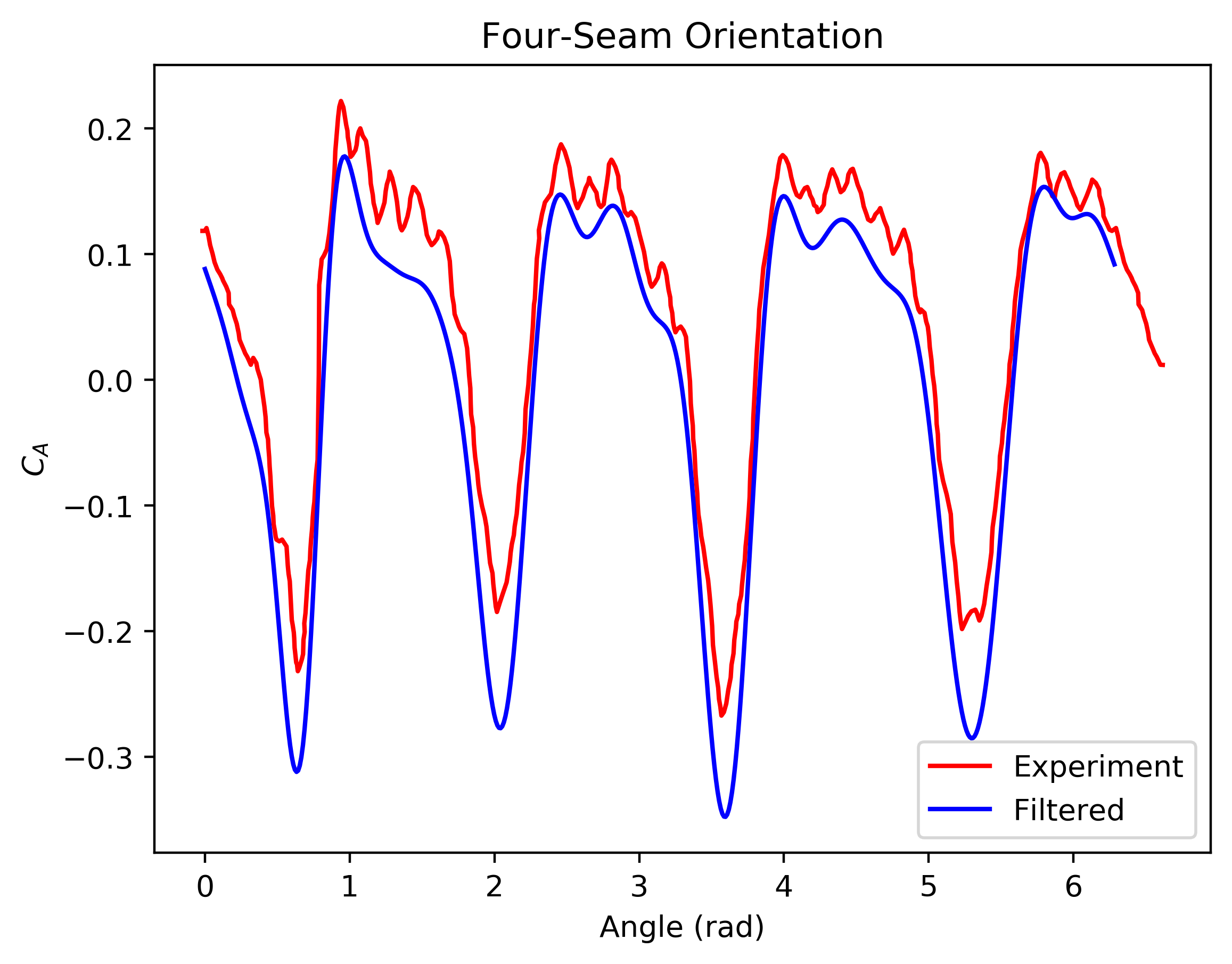}
\caption{\label{fig:data} The asymmetry force coefficient $C_A$ as a function of ball rotation angle in the two-seam (left) and four-seam (right) orientation. Shown are experimental data from Borg \& Morrissey \citep{Borg2014} (red line) and the filtered data (blue line) which is used in our Type C simulations. The filtered values show good overall agreement with the empirical data, while avoiding high wavenumber variations that can cause numerical artifacts in the differential equation solver.}
\end{figure*}

It is likely that the vertical and horizontal components of the asymmetry force both depend on the full angle vector, however as a simplification that will allow us to use the experimental data, we have chosen to consider models where the non-constant terms in the asymmetry force take the form 
\begin{equation}
    C_A \hat{\theta } \times \hat{v} =  G(\phi) \hat{x} + H( \psi ) \hat{z} \; ,
\end{equation}
where $\phi$ represents the angle about the ball's $\hat{z}$ axis, and $\psi$ similarly about the ball's $\hat{x}$ axis. This is a clear simplification as both $G$ and $H$ likely depend on the full angle vector $\vec{\theta}$, as well as on $\vec{\omega}$ and $\vec{v}$. Figure~\ref{fig:forces} shows a schematic diagram of the coordinate system and forces discussed above. 

If we combine Eqns.~\ref{eq:2nd}, \ref{eq:drag}, and \ref{eq:F_A} and separate our equations into components, we are left with
\begin{equation}
    \frac{ d u }{ d t} = - K_D u | \vec{v} | + K_A | \vec{v} |^2 G ( \phi )
    \label{eq:u}
\end{equation}
\begin{equation}
    \frac{ d v }{ d t} = - K_D v | \vec{v} | 
    \label{eq:v}
\end{equation}
\begin{equation}
    \frac{ d w }{ d t} = - K_D w | \vec{v} | + K_A | \vec{v} |^2 H ( \psi ) - g
    \label{eq:w}
\end{equation}
Here we have combined constants from previous equations so that only three remain $K_D$, $K_A$, and $g$, and defined the velocity vector $\vec{v} = \left[u, v, w \right]$. One obvious simplification of these equations can be achieved by setting $\phi = 0$. In this case, the dynamics of the problem will be confined to a single plane. This two-dimensional model will be confined to the $y$-$z$ plane if $u = 0$ initially. In this way we can explore the effect of the asymmetry term on only one direction of motion at a time. Our treatment of $\vec{F}_A$ has resulted in the introduction of two new quantities, namely the angles $\phi$ and $\psi$. For the case of a completely non-rotating ball, these are simply constants, however knuckleballs generally do show slow rates of rotation and experimental data has shown that the asymmetry force yields significant torque on the ball in flight \cite{Borg2014}. The drag force can also, in principle, exert a torque as well, however we will not consider it in this paper as the torque due to the drag force has not been experimentally considered independently from the torque due to the asymmetry force. 

\begin{table}[h!]
\centering
\begin{tabular}{lrr}
Variable & Value (Physical) & Value (Non-dimensional) \\
\hline
$x$ & 0.00 m & 0.00 \\
$y$ & 16.76 m & 1.00 \\
$z$ & 1.98 m & 0.12\\
$u$ & 0.0 m/s & 0.00\\
$v$ & -35.97 m/s & -1.07 \\
$w$ & 0.76 m/s& 0.02 \\
$\phi$ & 0 to $2 \pi$ rad & 0 to 1 \\
$\psi$ & 0 to $2 \pi$ rad & 0 to 1 \\
$ \omega_z$ & $-21.0$ to $21.0$ rad/s & -6.68 to 6.68 \\
$ \omega_x $ & $-21.0$ to $21.0$ rad/s & -6.68 to 6.68 \\
\end{tabular}
\caption{\label{tab:ICs} Initial conditions used for all simulations unless otherwise noted. }
\end{table}

In this paper we will consider three types of models, which we label Type A, Type B, and Type C. All three types solve Equations~\ref{eq:u}, \ref{eq:v}, and \ref{eq:w}, but they differ in their treatment of the rotation of the ball. Our Type A models assume constant angular velocities about each axis, so that $\vec{\theta} = \vec{\omega} t + \vec{\theta_0}$. Type A models essentially assume that any torque on the ball is insufficient to significantly alter the rotation of the ball in flight. 

The constant rotation models, however, ignore the torque associated with the force generated on the ball by the asymmetrical air flow around it. A force in the $+ \hat{x}$ direction should, for example, produce a corresponding torque on the ball in the $ \hat{z}$ direction unless it is acting precisely at the center of mass of the ball. If this force has any effective lever arm it would then produce a torque, leading to a change in the rotation rate. To date there is no available data on the changes in rotation rate experienced by a knuckleball in flight, so we will introduce an effective lever arm $\vec{\ell}_A$ for $\vec{F}_A$, such that the torque applied to the ball is simply given by 
\begin{equation}
    \vec{\tau}_A = \vec{\ell}_A \times \vec{F}_A = \ell_A F_{Az} \hat{x} - \ell_A F_{Ax} \hat{z} \; .
    \label{eq:torque}
\end{equation}
We chose to parameterize our effective lever arm in terms of the radius of the ball and a parameter $\alpha$, defined such that $-1 < \alpha < 1$, leading to $\ell_A = \alpha r$. Thus a choice of $\alpha = \pm 1$ would indicate the maximum possible lever arm, while a choice of $\alpha = 0$ would yield no torque. The evolution equations for the rotation of the ball about the $x$ and $z$ axes, respectively, are
\begin{equation}
    \frac{ d^2 \psi }{d t^2}  = - K_\psi \alpha | \vec{v} |^2 G(\psi) 
    \label{eq:psi}
\end{equation}
and
\begin{equation}
    \frac{ d^2 \phi }{d t^2} = K_\phi \alpha | \vec{v} |^2 H(\phi) \; .
    \label{eq:phi}
\end{equation}
Here again we have collected all numerical constants into $K_\psi$ and $K_\phi$. 

For Type A and B models, we choose a relatively simply functional form for our angular dependence where $G (\psi) = C_G \sin 4 \psi$ and $H (\phi) =  C_H \sin 4 \phi$. Figure~\ref{fig:data} shows the angular dependence of the asymmetry force, effectively showing $C_A$ for two orientations of a slowly rotating baseball. Examination of the data in Figure~\ref{fig:data} shows that both orientations produce variations that are roughly four-fold periodic over a rotation, and indeed a Fourier transform confirms that the four-fold periodic mode has the most power in both data sets. For Type A and B models we choose $C_G = C_H = 0.25$.

For Type C models we chose to employ the experimental data to create $G (\psi)$ and $H (\phi)$. Figure~\ref{fig:data} shows the experimental data, as well as the smoothed, filtered version of the same data that we employ in our models. We have interpolated the experimental data onto a regular grid in angle and then applied a Gaussian smoothing to avoid small-scale features which will likely produce unreliable numerical results. Additionally, we have subtracted the mean value from both datasets as a mean value would imply a preferred direction that would follow the coordinate system under a rotation of the axes. These data are then used to generate a cubic interpolating polynomial to provide an empirical function for our differential equation solver.

In summary, we have three groups of models to explore. They are:
\begin{itemize}
    \item Type A: Constant rotation
    \item Type B: Torque feedback with $\sin{4 \psi}$ asymmetry force
    \item Type C: Torque feedback with empirical asymmetry force
\end{itemize}
For each group of models, we can either consider 2D motion with rotation about a single axis, or 3D motion with rotation about two axes. 

To solve these equations we employ the SciPy differential equation solving library \cite{Community2019}. Specifically, we have use an adaptive 4th/5th order Runge-Kutta solver \cite{Dormand1980} to solve Eqns.~\ref{eq:u}, \ref{eq:v}, \ref{eq:w}, \ref{eq:psi}, and \ref{eq:phi}. This implementation includes the ability to provide a maximum level of relative numerical error per timestep.  Initial conditions were selected to be realistic for a knuckleball thrown by a professional pitcher, and are given in Table~\ref{tab:ICs}. For the initial orientation of the ball ($\psi$ and $\phi$) and the initial angular velocity components ($\omega_x$ and $\omega_z$) a range of values are considered. 

\section{Measuring Chaos in Our Models}

\begin{figure}[t]
\centering
\includegraphics[width=0.6\linewidth]{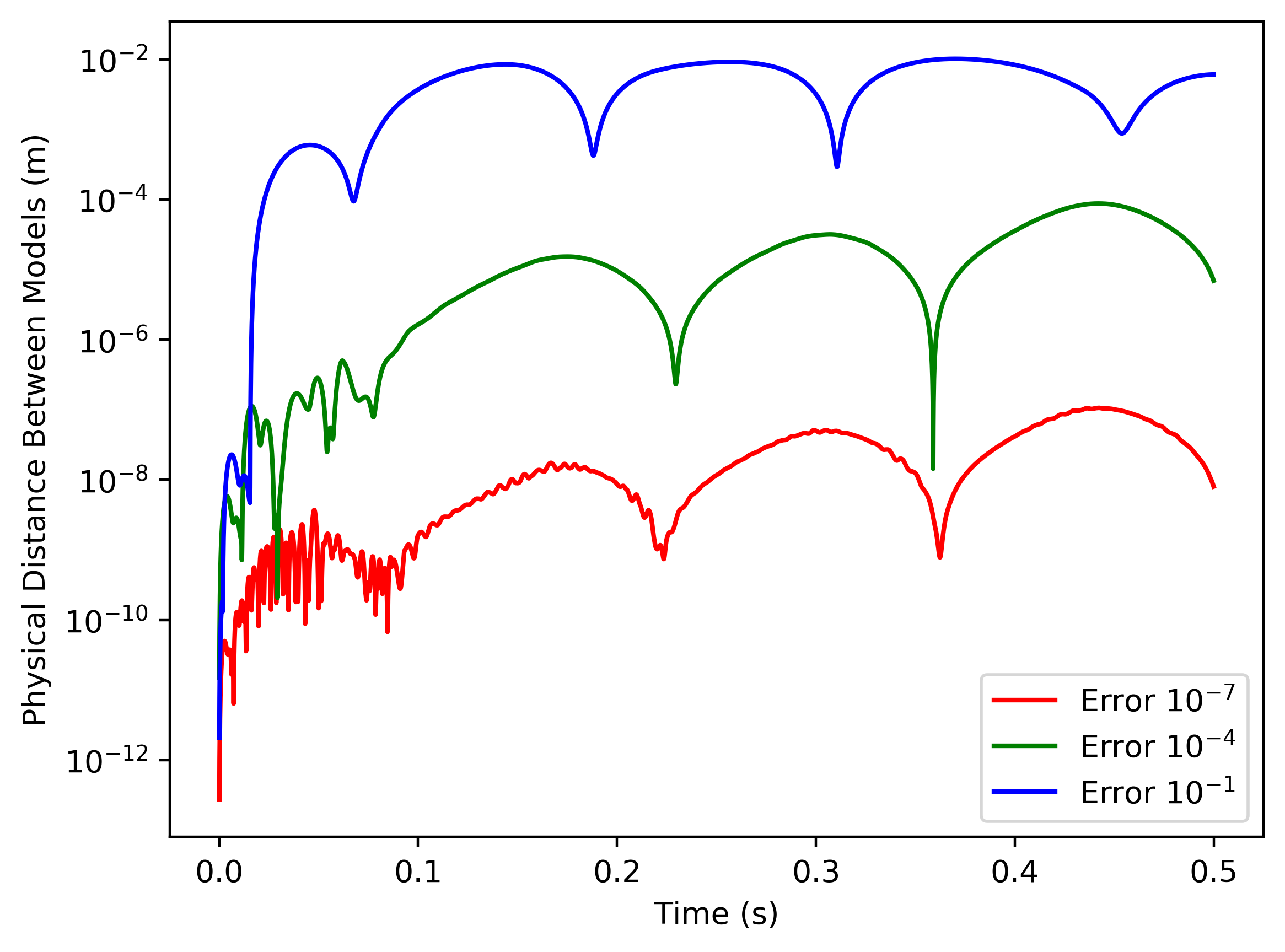}
\caption{\label{fig:convergence} Calculated physical distance between models with identical initial conditions but varying levels of relative error per timestep. Distances are calculated compared to a model with a relative error of $10^{-10}$ per timestep. This figure shows that in order to assure a maximum accumulated numerical error of less than one part in $10^{6}$, we must employ a relative error per timestep of no more than $10^{-7}$. }
\end{figure}

Our three models for the trajectory of a knuckleball are prime candidates for the investigation of dynamical chaos as they are strongly non-linear in both the drag and asymmetry forces. These are deterministic equations, so if we provide suitable initial conditions the trajectory of the ball should be calculable for all future time given sufficient numerical accuracy in our method of solution, and indeed numerical experiments show this to be the case. Systems like our knuckleball models, however, can exhibit dynamical chaos and thus show extreme sensitivity to their initial conditions. In our case, that could mean that changing the initial orientation or angular velocity of the ball by an infinitesimal amount may yield large, qualitative differences in outcome when the ball reaches the plate. It is also important to note here that unlike many prototypical chaotic systems like the damped driven pendulum or the double pendulum, a knuckleball does not exhibit periodic motion. This limits the conceptual tools available to measure chaotic behaviors.

\begin{figure*}[t]
\includegraphics[width=0.49\linewidth]{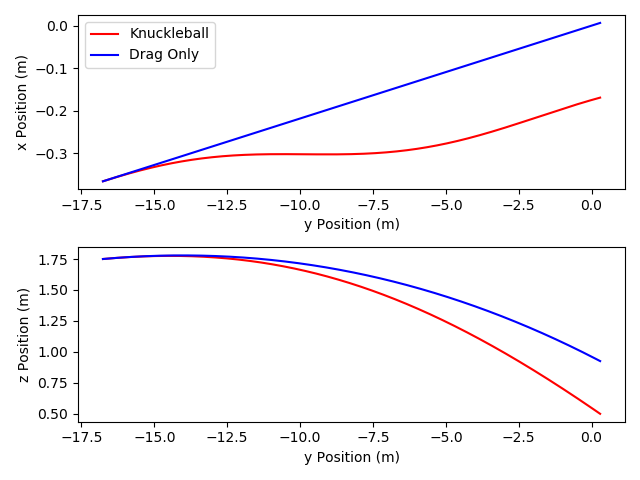}
\includegraphics[width=0.49\linewidth]{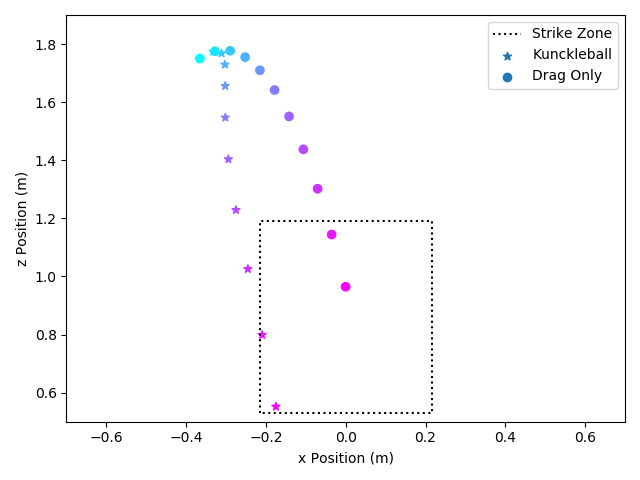}
\caption{\label{fig:traj_ConstRot} Trajectory of a Type A knuckleball with initial conditions $\psi (0) = 0.106$ rad, $\phi(0) = 0.114$ rad,  $\omega_x (0) = -0.506$ rad/s, and $\omega_z (0) = -2.001$ rad/s, shown as a function of $y$ position (left panel, red line) as well as snapshots of the location on the $x$-$z$ plane from the batter's viewpoint (right panel, stars) where the color of the marker indicates the distance from the plate ranging from the pitcher (light blue) to arrival at the plate (pink). Also shown for comparison is the trajectory of the same pitch without the effects of the asymmetry force (blue line on left, circle markers on right), and the location of the strike zone (dotted line). Thus even the simplest model considered can show behaviors that qualitatively match those seen in real knuckleballs.}
\end{figure*}

In order to search for chaos, we need to consider the phase space of these models. Type A models have a three-dimensional phase space, while Type B and C models have a seven-dimensional phase space.    To examine the behavior of these trajectories through phase space, it is necessary to scale our physical quantities and hence non-dimensionalize them. Here we choose the initial $y$ position of the ball as our length scale, the approximate travel time of the ball to home plate of 0.5 s as our timescale, and $2 \pi$ as our angular scale. We will use these scale factors in all discussions of phase space distance. Table~\ref{tab:ICs} also shows the non-dimensional values for our initial conditions. We will denote non-dimensional quantities with tildes.

In order to investigate the presence of dynamical chaos in these models, we need to calculate the distance in phase space between solutions with differing initial conditions. The phase space vector for the most general case is given by $ \vec{\chi} =  \left[ \tilde{u}, \tilde{v}, \tilde{w}, \tilde{\psi}, \tilde{\omega}_x, \tilde{\phi}, \tilde{\omega}_z \right]$. We can define the distance in phase space between two models with different initial conditions as the Euclidean distance (e.g., distance $d$ is defined by $d^2 = \Delta x^2 + \Delta y^2$ for an $x$-$y$ plane) in that space, which we denote $ \Delta \chi $. To study potential chaotic behavior, we must be able to evolve two sets of trajectories through phase space which begin very close to each other. We have chosen initial perturbations on the order of $10^{-6}$. 

When searching for chaos in a system with a high-dimensional phase space and non-periodic motion through that phase space, the best mathematical tool for finding and quantifying chaos is the Lyapunov exponent. Lyapunov exponents measure the rate at which two trajectories through phase space converge or diverge as they are evolved through time. Dynamical systems such as our models for the knuckleball can be described by a set of Lyapunov exponents, which describe the local convergence or divergence of a set of trajectories through phase space in each dimensional of that phase space. Chaotic systems are those in which at least one of the Lyapunov exponents of the system is positive for a given choice of parameters \cite{Strogatz2015, Brown2018}. In practice, the largest Lyapunov exponent can be measured by determining the exponential growth rate of $\Delta \chi$ over a given time interval. Lyapunov exponents have units of s$^{-1}$, and thus represent the number of factors of $e$ that two slightly different trajectories in phase space will diverge or converge by per second. 

Perhaps more practically, they indicate the intrinsic growth of uncertainty in the location of the model in phase space. As a simple example, if we assume that the phase space location of a ball leaving the pitcher's hand could be measured to 0.1\% accuracy in each component of the ball's velocity, orientation, and angular velocity and that choice of initial conditions yielded a set of Lyapunov exponents which all had values of 10 s$^{-1}$, then when the ball crossed home plate the uncertainty may have grown to roughly 15\% in each component. In practice, the uncertainty on all components will not grow equally, but as we will show it is not unreasonable to obtain many $e$-fold increases in phase space divergence over the course of a chaotic pitch.

It is also important to note that in almost all chaotic systems there are physical limits beyond which two systems cannot diverge. These limits are often related to conserved quantities such as mass, momentum, or energy. Even in non-conservative systems such as the knuckleball, the change in these quantities is limited by the magnitude of the forces, torques, and/or work on the system. This can be referred to as nonlinear saturation of the exponential growth, as it is reminiscent of the nonlinear saturation of instabilities in fluid mechanics. 

\begin{figure*}[t!]
\centering
\includegraphics[width=0.49\linewidth]{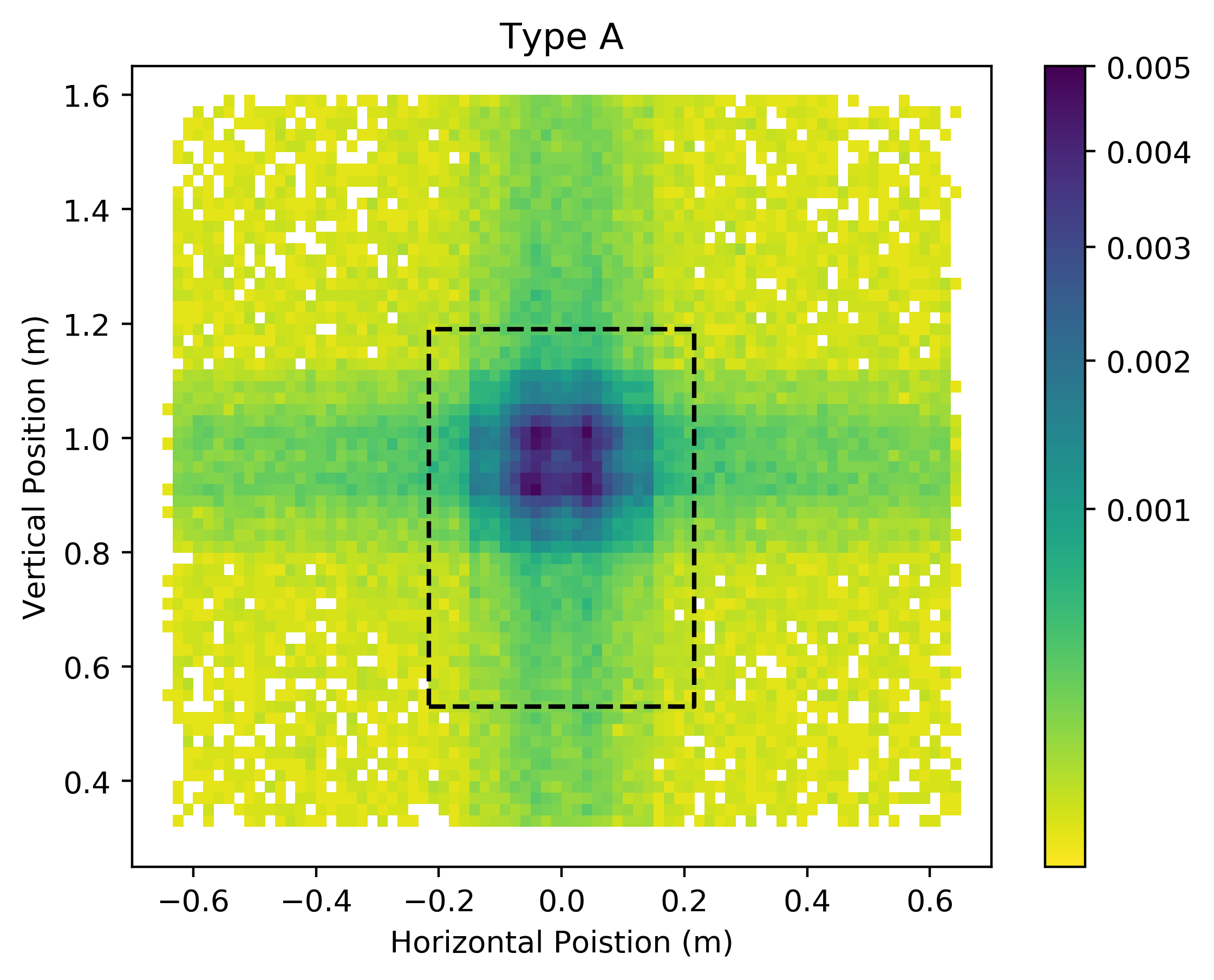}
\includegraphics[width=0.49\linewidth]{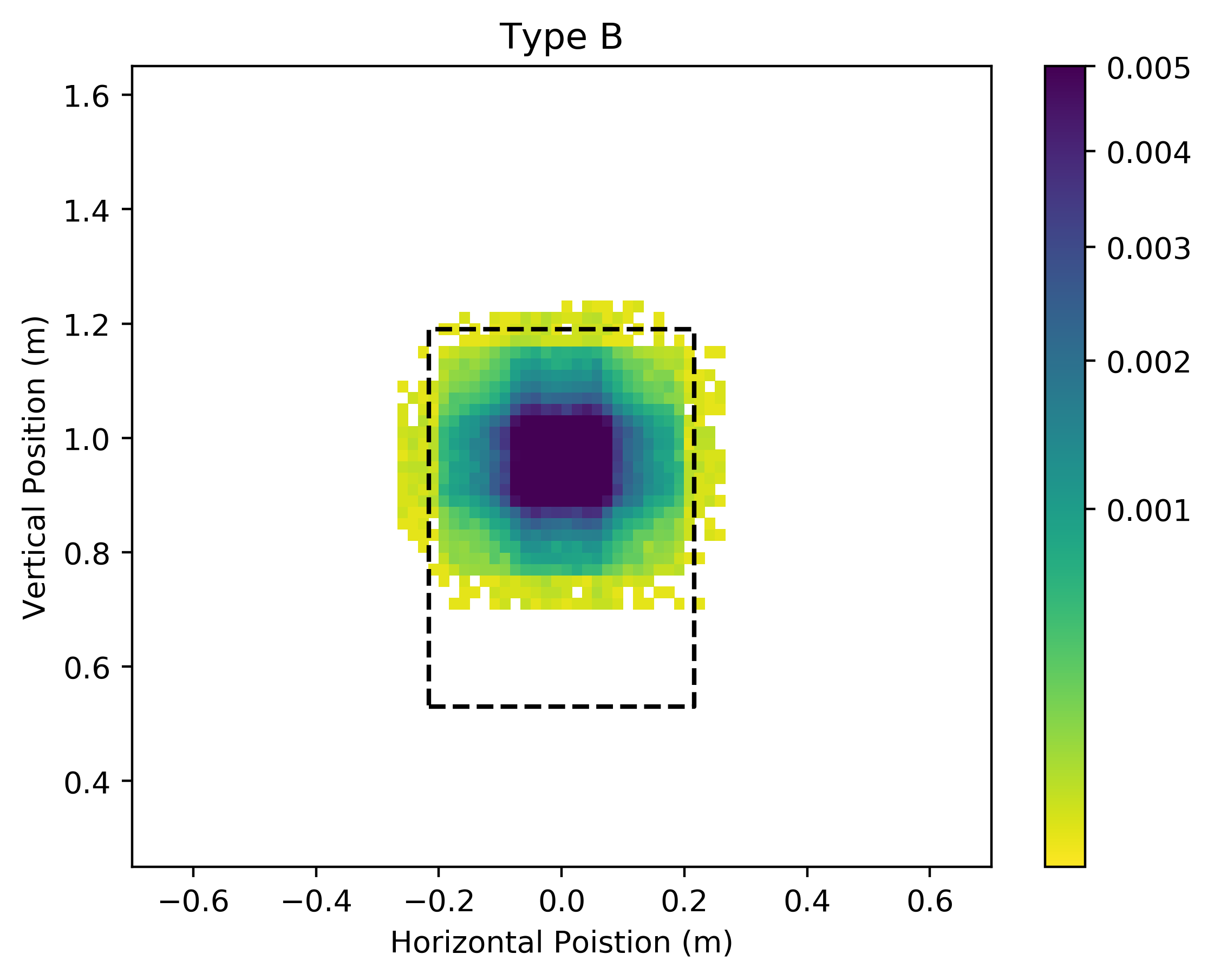}
\includegraphics[width=0.49\linewidth]{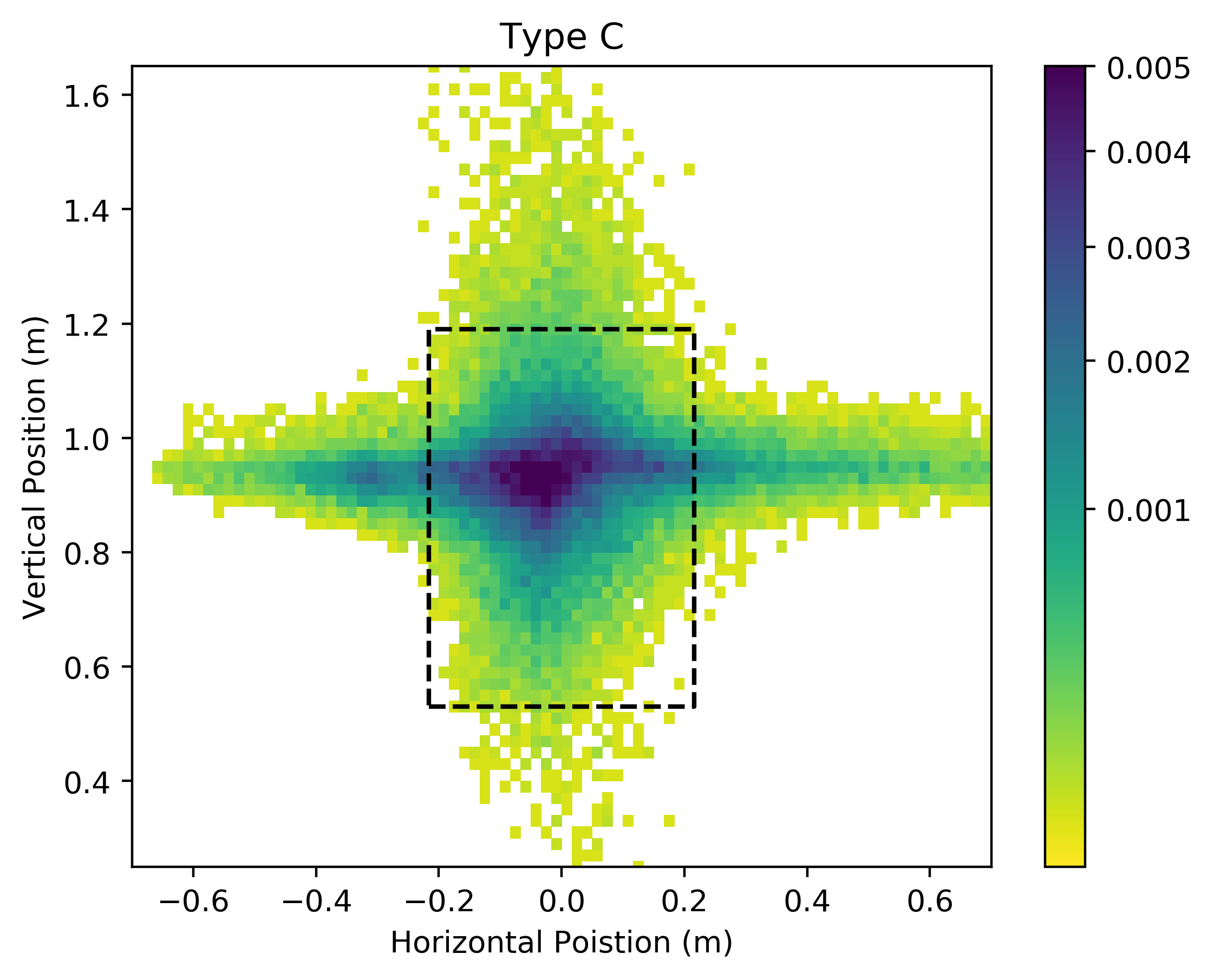}
\includegraphics[width=0.49\linewidth]{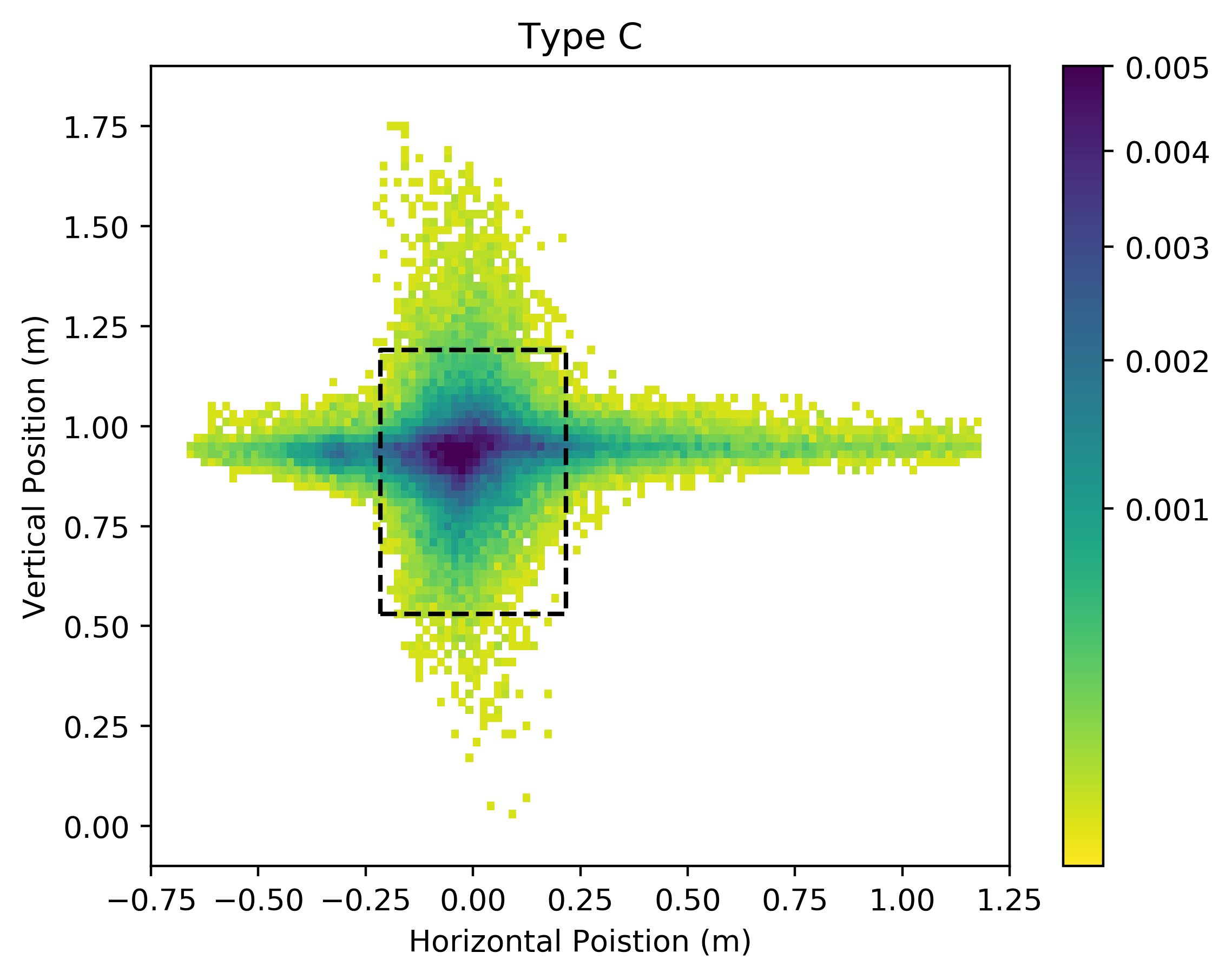}
\caption{\label{fig:PlateLocs_2dHist} Two-dimensional Histograms of the location of simulated knuckleballs with the initial conditions given in Table~\ref{tab:ICs} at home plate for Type A (upper left), Type B (upper right), and Type C models (bottom row). Also shown is the strike zone (black dashed line) for reference. For Type C we have shown the same view as for Types A and B (lower left), as well as a zoomed-out view (lower right) that shows all simulated plate locations. Types B and C used $\alpha = 0.5$. Initial angles were randomly selected from a uniform distribution between $-\pi/4$ and $\pi/4$ and initial angular velocities were randomly selected from a Gaussian distribution centered at 0 rad/s with a width of 8 rad/s. Here dark blue indicates large fractions of pitches ended up at that location, while light green indicates very few pitches. Note that the color scale here uses a power-law to highlight the regions with very few pitches. All three model types show substantial variation in plate location, as expected from knuckleballs. }
\end{figure*}

To investigate the exponential growth of uncertainty in these models, we will need to control numerical error so that it is always much smaller than the phase space separation between two models we wish to compare. In order to access the exponential separation To investigate the accuracy of our numerical solver, we conducted a convergence test, comparing the time evolution of a single set of initial conditions at various levels of relative error per timestep.  Figure~\ref{fig:convergence} shows the physical space distance between a reference solution computed with a relative error per timestep of $10^{-10}$. Given this evolution of numerical error, we choose to conduct all models with a relative numerical error of $10^{-7}$ per timestep and use initial phase space distances between pairs of initial conditions of $10^{-6}$. This assures that our numerical error will remain much smaller than the phase space distances we desire to measure.

\section{Examples of Knuckleball Trajectories}

Having developed a general framework for our three types of knuckleball models, let us first explore if the trajectories they produce show qualitative agreement with observed properties of the knuckleball -- namely that they show substantial movement in flight and that they work best for very low levels of initial angular velocity. 

We will first consider the behavior of models with constant rotation, which we label Type A. For these models the angular velocity vector is constant, resulting in a periodic force on the ball as it travels towards home plate. Though the amplitude of the periodic forcing is proportional to the velocity squared, the dominant component of velocity in the $y$-direction means that the periodic movement of the ball in the $x$- or $z$-directions has little effect on the magnitude of the velocity and hence the amplitude of the forcing. These models do, however, show substantial effects similar to those seen in real knuckleballs. 

Figure~\ref{fig:traj_ConstRot} shows the trajectory of a sample knuckleball with constant rotation, compared with the trajectory of a an identical pitch with no asymmetry force. This Type A model reproduces a key feature of knuckleballs, specifically the ability to change directions in mid-flight. In this example, the knuckleball experiences a strong leftward acceleration initially which not only stops the ball's rightward motion but reverses it.  The acceleration then changes sign again, leading to a ball which was moving away from the strike zone to instead just slip in the lefthand edge. Similarly, this pitch shows roughly only gravitational acceleration downward to start, but as the ball approaches the plate its downward acceleration increases. Thus what looks initially like a pitch in the upper center of the strike zone changes through the course of its flight to look as if it will be high and left of the strike zone. Finally, this pitch turns back to the right and accelerates downward and ends up catching the lower left corner of the strike zone. It should be re-emphasized that this knuckleball was thrown with identical initial position and velocity to the drag-only pitch. Similar antics are seen by knuckleballs with a variety of initial orientations and rotation rates for Type A, B, and C models.

\begin{figure}[t]
\centering
\includegraphics[width=0.6\linewidth]{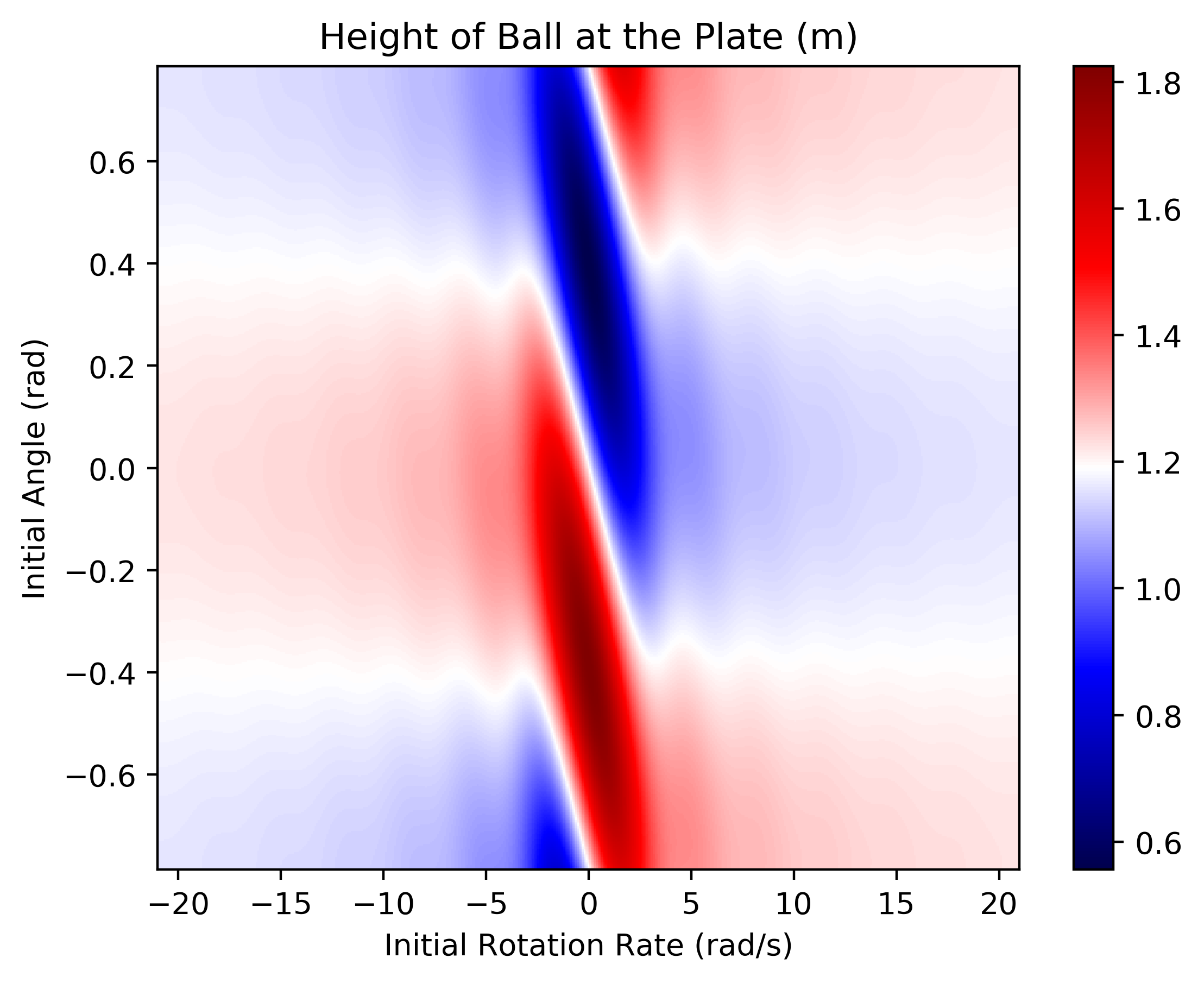}
\caption{Height of the ball at the plate as determined by changing the initial angle $\psi$ and the angular velocity $\omega_x$ for Type A knuckleball models with constant angular velocity. All pitches have identical initial position and velocity. Knuckleballs show up to 60 cm of movement up (red) or down (blue) due to the asymmetry force when slowly rotating. Thus even the most simple model considered shows substantial variation due to the initial orientation and spin of the ball. \label{fig:noTorque_PlateLocs} }
\end{figure}

Perhaps the most obvious question to examine with the three knuckleball models to to ask what range of locations as the ball crosses the plate can be achieved. In order to investigate this we used the same initial position and velocity shown in Figure~\ref{fig:traj_ConstRot} and ran $10^5$ simulations for random initial angles between $-\pi / 4$ and $\pi / 4$ and random initial angular velocities taken from a normal distribution centered on zero with a width of 8 rad/s. 

Figure~\ref{fig:PlateLocs_2dHist} shows a 2D histogram of the locations $10^5$ simulated pitches arriving at home plate in the $x$-$z$ plane for Type A (upper left), Type B (upper right), and Type C (lower) models. The histograms have been normalized so that the color represents the fraction of pitches per bin. All three types show the possibility of substantial movement relative to a drag-only pitch, though the amount of motion possible varies significantly. Type A simulations show as far as 70 cm of movement in either the $x$- or $z$-directions, leading to box shaped region with a cross in the center. Type B models show a similar box and cross pattern, but it is much smaller, with only about 25 cm of movement possible. Type C models exhibit only the cross and show much more potential motion, with pitches moving as much as 1.2 m horizontally and 80 cm vertically. This indicates that for Type C pitches there is significant interaction between the two axes of motion.

\begin{figure*}[t]
\includegraphics[width=0.49\linewidth]{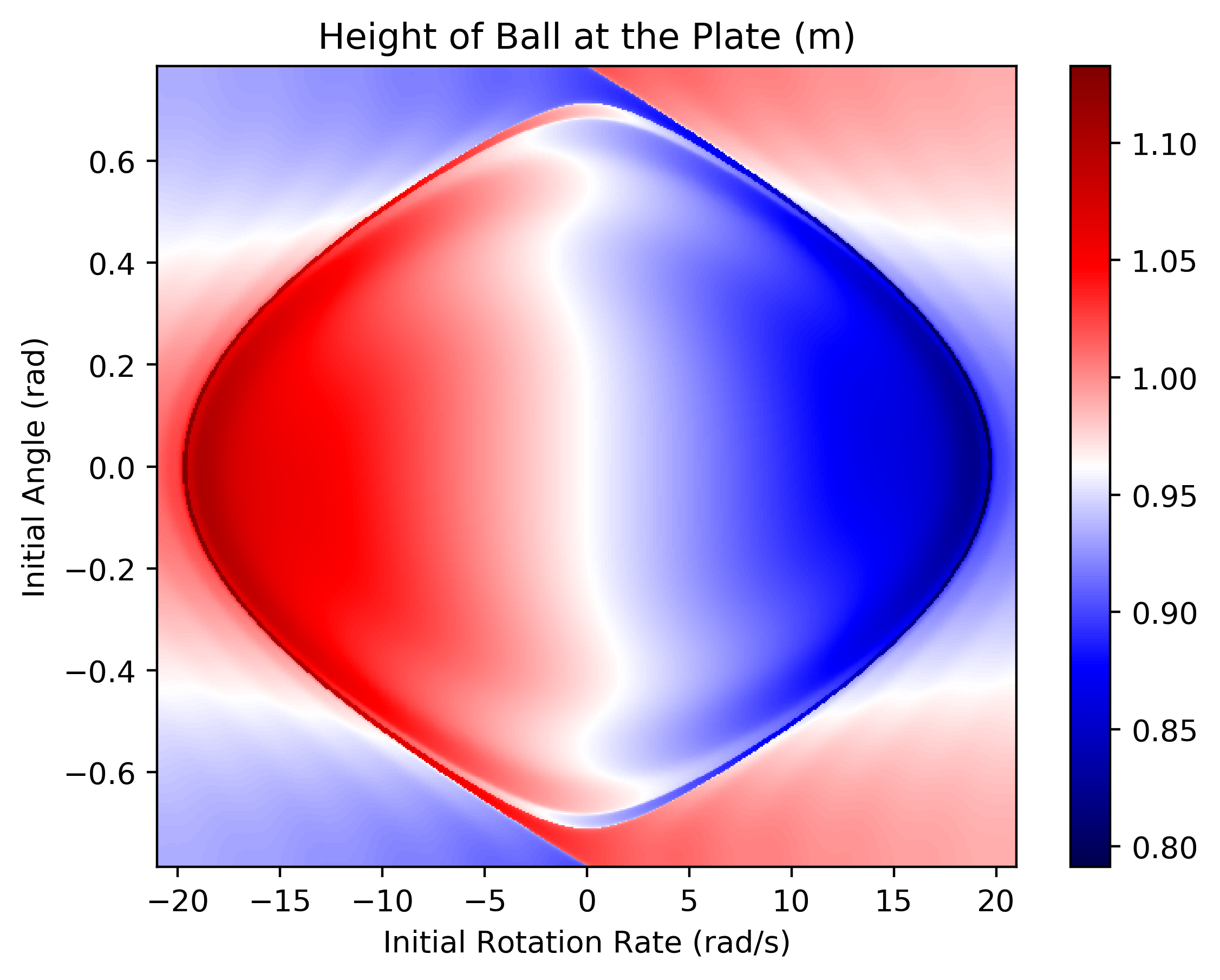}
\includegraphics[width=0.49\linewidth]{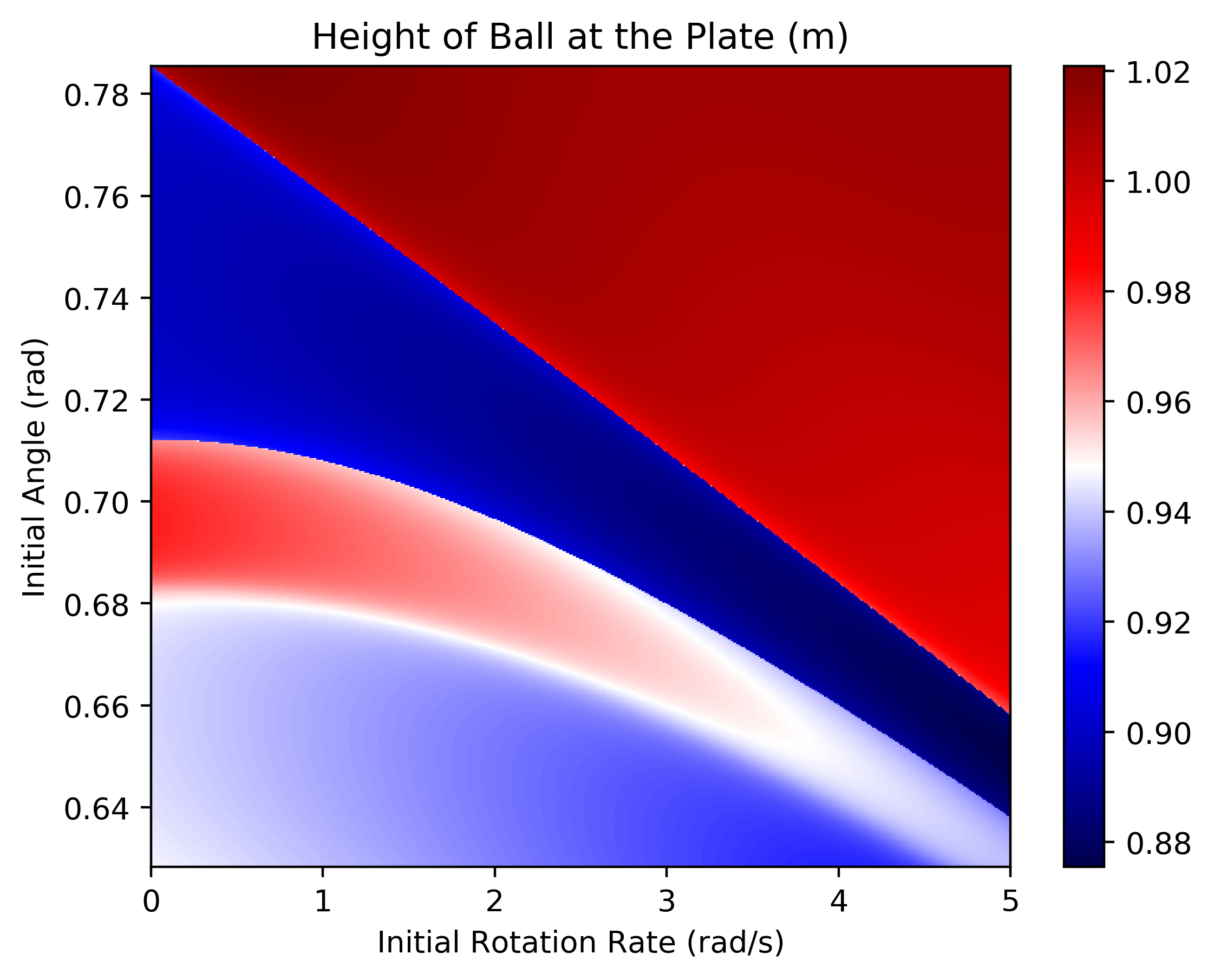}
\caption{\label{fig:Torque_PlaceLocs} Left: Height of the ball at the plate for Type B model pitches the same set of initial conditions used in Figure~\ref{fig:noTorque_PlateLocs} and with $\alpha = 1$. A drag-only pitch would arrive at the plate with a height of 0.96 m (white), while knuckleballs are simulated to vary between 0.79 m (dark blue) and 1.14 m (dark red). Right: Zoom-in of right hand panel highlighting the same region in parameter space shown in right-hand panel of Fig.~\ref{fig:PhaseSepMap}. The sharp transitions between pitches going high (dark red) and low (dark blue) correspond to the arcs of chaotic behavior seen in Fig.~\ref{fig:PhaseSepMap}. In addition to substantial variation in trajectory with initial orientation and spin, these sharp transitions also hint at the possibility of bifurcation surfaces in phase space. }
\end{figure*}

To further explore the range of possible behaviors for our Type A, B, and C models, we conducted a systematic exploration for a set of 2D models where motion is confined to the $y$-$z$ plane and only rotation about the $x$-axis is considered. We model the trajectories of pitches with $-\pi/4 < \psi(0) < \pi/4$ rad and $-21.0 < \omega_x (0) < 21.0$ rad/s, and measure the height of the ball when it reached the plate.

Figure~\ref{fig:noTorque_PlateLocs} shows the height of the ball at the plate for a range of initial angles and angular velocities for Type A models. By achieving very low rotation rates the pitch can achieve very large changes in its height at the plate simply by changing the initial orientation of the ball. Interestingly, the total deviation of the ball from its drag-only trajectory appears to be greatly reduced as angular velocity increases in these models, with the largest reductions in the deviation from the drag-only trajectory for angular velocities less than about 3 rad/s or roughly one-quarter revolution over the half-second in which the ball is in flight. This is qualitatively in agreement with how pitchers are taught to throw knuckleballs with ``less than a half turn of the ball on its way to the plate'' as ideal for generating large and erratic motions \cite{Clark2012}.

 Figure~\ref{fig:Torque_PlaceLocs} shows the height of the ball at the plate for the Type B models over the same range of initial conditions. Our simulated trajectories move the ball as high as 1.14 m and as low as 0.79 m compared to a drag-only trajectory, which would cross the plate at a height of 0.96 m. Perhaps most striking are the sharp changes in height that are observed for initial angles near $\pm \pi/4$ and angular velocities near zero. The right-hand plot in Figure~\ref{fig:Torque_PlaceLocs} shows a zoomed in region to highlight the sharpness of the jump between pitches that arrive high (red) and low (blue). The sharp transitions indicate that two pitches thrown with nearly identical initial conditions end up at very different heights when they reach the plate. This is tantalizing evidence for dynamical chaos in our Type B models, as they appear to be bifurcation surfaces where random perturbations can move the trajectory towards different basins of attraction, namely arriving at the plate higher or lower than expected.

\begin{figure}[t]
\centering
\includegraphics[width=0.6\linewidth]{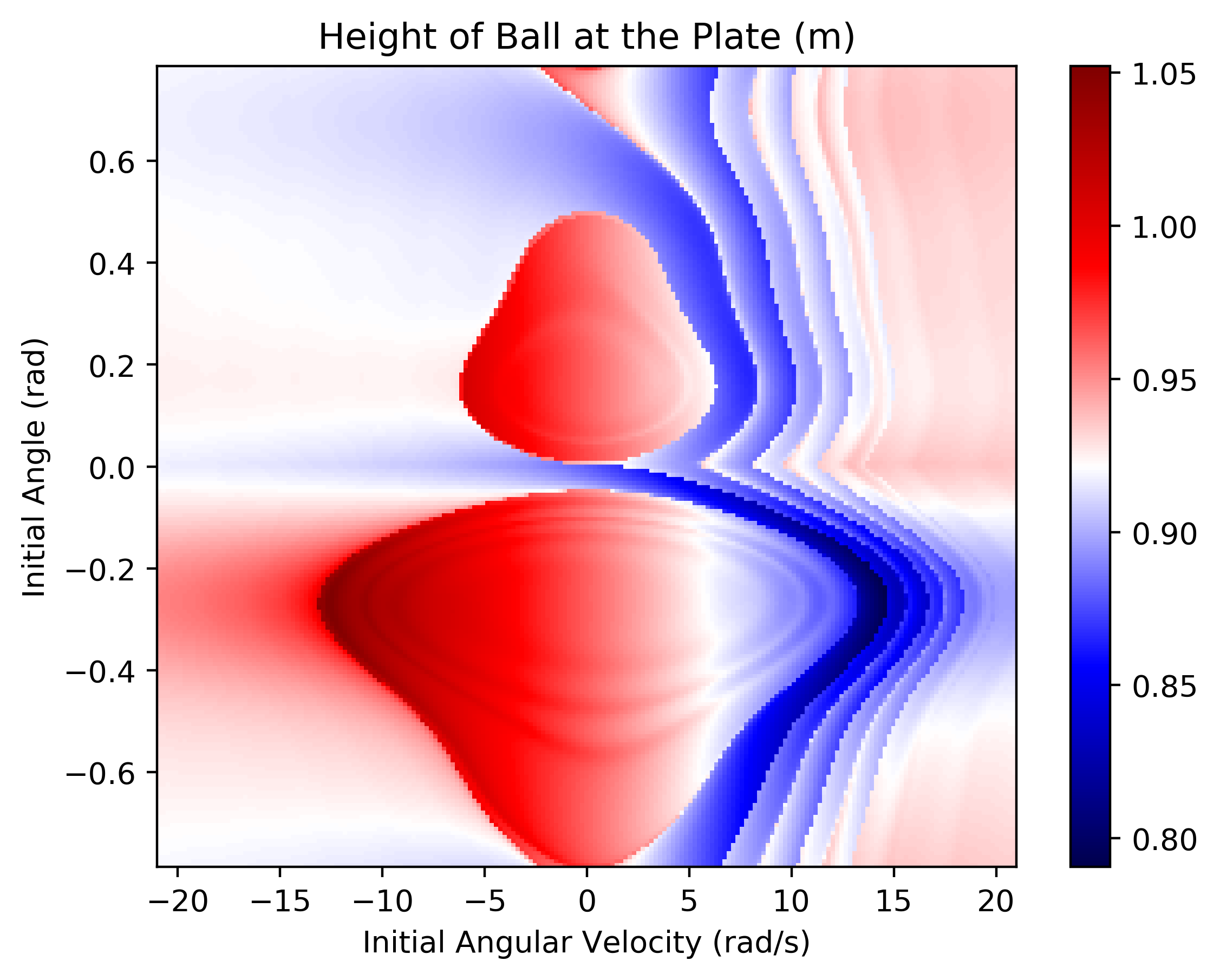}
\caption{\label{fig:PlateLocs_realCL} Height of the ball at the plate for Type C model pitches in the two-seam configuration with the same set of initial conditions described in Fig.~\ref{fig:Torque_PlaceLocs}. A drag-only pitch would arrive at the plate with a height of 0.96 m (bright red), while knuckleballs are simulated to vary between 0.78 m (dark blue) and 1.06 m (dark red). Here many potential bifurcation surfaces are seen, providing hints of chaotic behavior.}
\end{figure}

Figure~\ref{fig:PlateLocs_realCL} shows the height of a pitch at the plate with otherwise identical initial conditions but varying the initial orientation and angular velocity for Type C models. As in previous cases, a pitch with only drag forces and gravity would arrive at the plate with a height of 0.96 m (light red). This may seem to conflict with the wider range of positions seen in Figure~\ref{fig:PlateLocs_2dHist} for Type C models, however this may simply be a statement of the rarity of the extreme values. Additionally the extreme values may be a product of the additional complexity when 3D motion and two-axis rotation is considered.

Type C models tend to produce more trajectories lower than the drag-only case than they do higher trajectories at the plate. Interestingly, pitch tracking data shows a similar preference for downward motion \cite{Nathan2012}. As might be expected given the additional complexity of the Type C models, Figure~\ref{fig:PlateLocs_realCL} shows rich, complex structure with many bifurcation surfaces. These sharp changes in pitch behavior illustrate why predicting the path of the ball might be seemingly impossible, and they provide tantalizing hints of chaotic behavior. 


\section{\label{sec:chaos}Chaos Depends on Torque}

While all three types of knuckleball models yield substantial variation in the location at which they arrive at the plate, the question of if they are chaotic still remains. For this we need to compute Lyapunov exponents and look for models where the distance between two nearly identical sets of initial conditions shows exponential growth in time. 

To test for chaotic behavior, we compute a second trajectory with initial conditions that have been perturbed by a very small amount. We apply a random perturbation to the initial angles and angular velocity components such that the initial phase space separation between the two initial conditions was $10^{-6}$ and then tracked the separation between the two trajectories in phase space. This is repeated for $10^6$ trajectories with initial angles and angular velocities randomly selected from uniform distributions. The initial angles are selected from a uniform random distribution between $-\pi/4$ and $\pi/4$ rad for both $\psi$ and $\phi$. The angular velocity components are also selected from a uniform random distribution between $-8.0$ and $8.0$ rad/s for both $\omega_x$ and $\omega_z$. This process is conducted for all three model types.

For Type A models none of the $10^6$ trajectory pairs computed show exponential separation. All models begin with $\Delta \chi (0) = 10^{-6}$ and at home plate none of the phase space separations grow by more than a factor of four. Thus Type A models do not show dynamical chaos for the range of parameter space covering baseball pitches. 

\begin{figure}[t]
\centering
\includegraphics[width=0.6\linewidth]{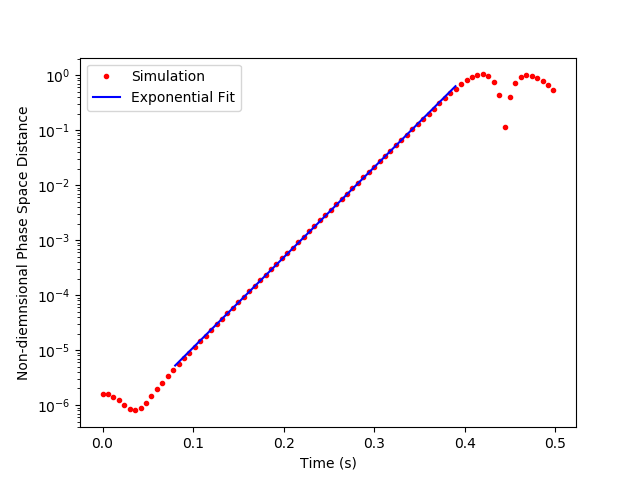}
\caption{\label{fig:phaseSep} Separation in phase space of two trajectories with nearly-identical initial conditions. The non-dimensional phase space distance $\Delta \chi$ grows exponentially from its initial value of approximately $10^{-6}$ through six orders of magnitude within 0.4 s, for a maximum Lyapunov exponent of 37.8 s$^{-1}$. The exponential growth of the phase space separation shows clear evidence of dynamical chaos.}
\end{figure}

\begin{figure*}[t]
\includegraphics[width=0.49\linewidth]{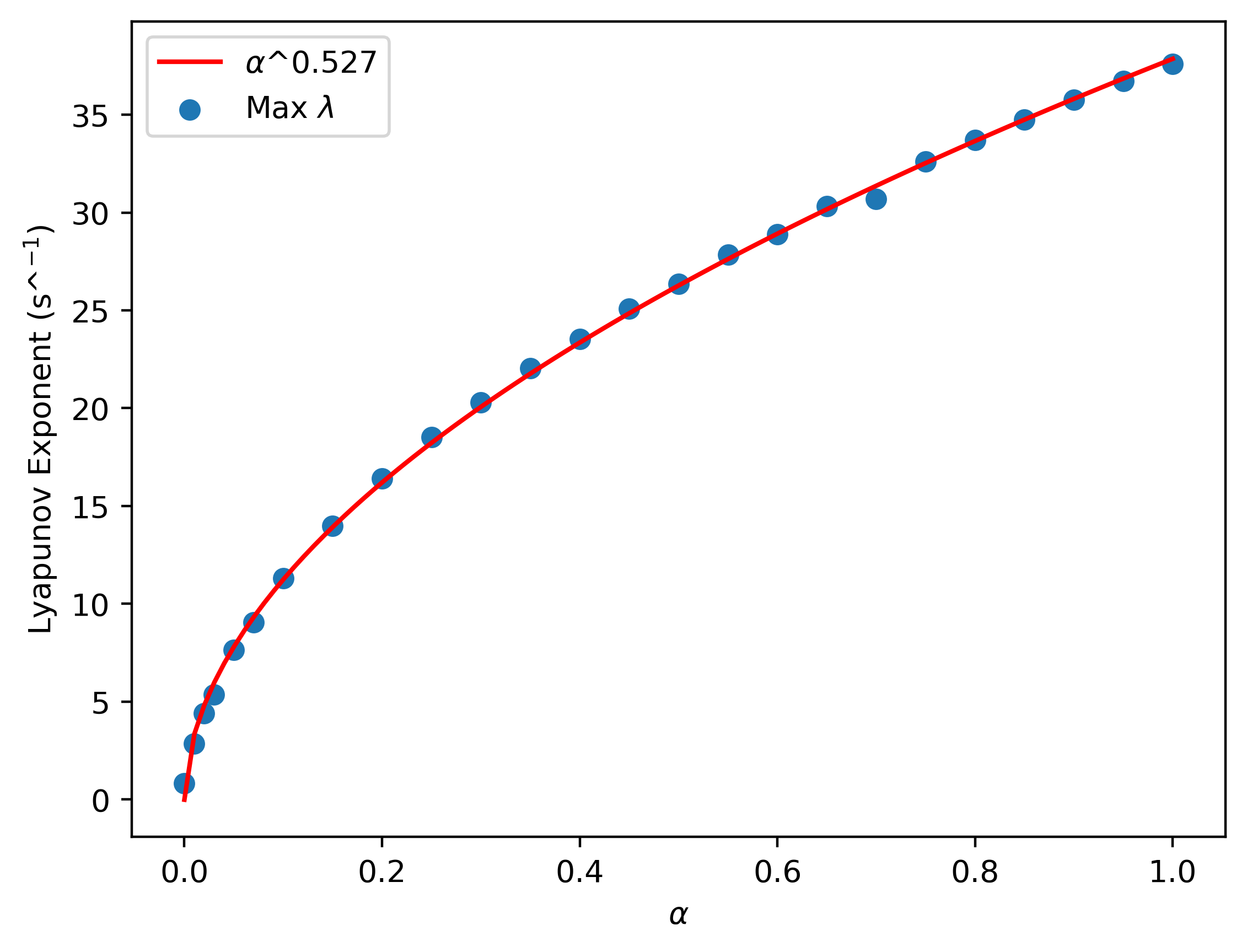}
\includegraphics[width=0.49\linewidth]{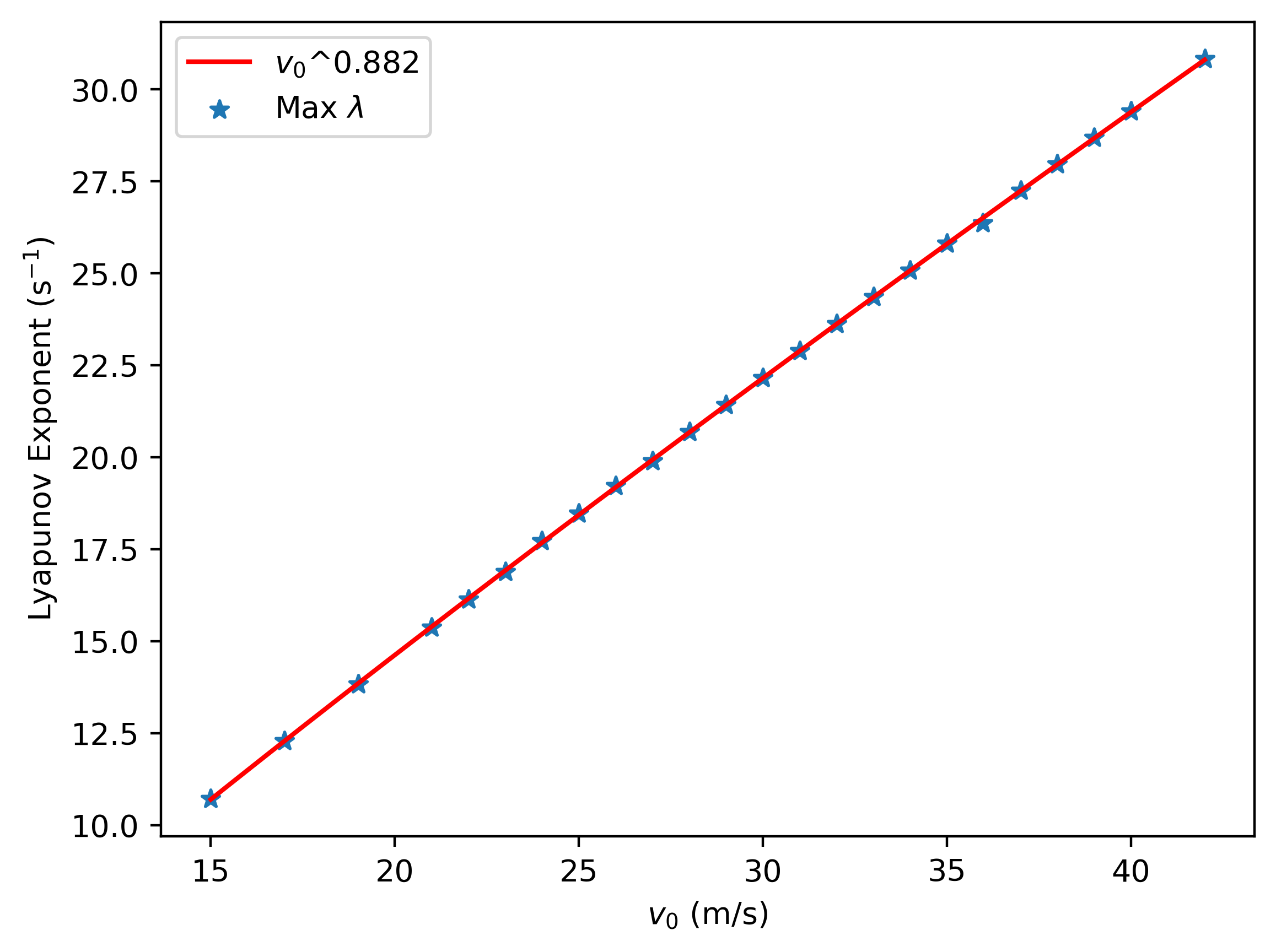}
\caption{\label{fig:lambdaDep} Largest Lyapunov exponent measured as a function of $\alpha$ (left panel) and as a function of $v_0$ (right panel). Largest Lyapunov exponents are calculated from $10^5$ trajectories with initial conditions $\pi/5 \leq \psi \leq \pi/4$ rad and $0 \leq \omega_x \leq 6.0$ rad/s for given values of the effective lever arm parameter $\alpha$  and the initial $y$ component of velocity $v_0$. Left Panel: Variation in the largest Lyapunov exponent achieved for a parameter sweep when $v_0 = 36$ m/s and $0 \leq \alpha \leq 1$ (blue dots).  Also plotted is the best fit power-law to the data that shows $\lambda \propto \alpha^{0.527}$ (red line). Right Panel: Same as left panel, but here $alpha$ is held constant at 0.5, while $v_0$ is varied between 15.0 m/s and 42.0 m/s (blue stars). The best fit power law (red line) shows $\lambda \propto v_0^{0.882}$. This shows how the free parameter $\alpha$ and the initial speed of the pitch $v_0$ impact the growth rate for chaotic uncertainty in the phase space location of the ball.}
\end{figure*}

Type B models show hints of chaotic dynamics in the bifurcation surfaces and in fact our search for dynamical chaos shows that Type B models can produce chaos. After computing $10^6$ trajectories for randomly chosen initial angles and angular velocities, we find that most trajectory pairs do not show exponential growth in $\Delta \chi$, however approximately 1.6\% do show at lest two orders of magnitude increase in phase space separation by the time they arrive at home plate. Thus while most choices of initial conditions do not yield chaotic behavior for Type B models, a significant fraction do.

The largest such separation in phase space is plotted in Figure~\ref{fig:phaseSep}. This trajectory has initial angles $\psi(0) = 0.696$ rad and $\phi (0) = -0.474$ rad, and initial angular velocity components $\omega_x (0) = 3.543$ rad/s and $\omega_z (0) = -2.413$ rad/s. Thus the pitch makes roughly one-quarter of a revolution about both axes on its way to the plate. A second trajectory with a random perturbation with a magnitude of order $10^{-6}$ was also computed, and the phase space separation grew more than six orders of magnitude as the ball travels towards the plate. This yields a maximum Lyapunov exponent of $37.8$ s$^{-1}$.  This indicates that the phase space separation between two trajectories will increase by a factor of $e$ roughly every 26 ms for this pitch. Due to the large positive value of the maximum Lyapunov exponent, this model with these initial conditions clearly demonstrates dynamical chaos.

\subsection{Effects of Lever-Arm and Pitch Speed}

Now that we have unambiguously found a chaotic set of parameters and initial conditions, we can explore how the behavior is changed by changing the effective lever arm of the feedback torque and the initial velocity of the pitch. One might reasonably expect that chaotic behavior will become more common for faster pitches and similarly for greater values of $\alpha$.

Our lever-arm parameter is of particular interest as it is essentially a free-parameter in our Type B and C models. It is difficult to estimate what values of $\alpha$ might be expected. Physical scenarios can be envisioned where $\alpha$ might range from -1 (all of the transverse force is applied at the leading edge of the ball) to 0 (the transverse force is applied uniformly to all points on one hemisphere of the ball) to 1 (all of the transverse force is applied at the trailing edge of the ball). Given the boundary layer separation which drives the motion \cite{Morrissey2009} we might reasonably expect that $\alpha > 0$, and if the transverse force is uniformly applied to the back quarter of the ball then $\alpha = 0.5$. Borg \& Morrissey report that in wind tunnels data they observe torques of  approximately $4 \times 10^{-3}$ N m, which would correspond to $\alpha \approx 0.3$ \cite{Borg2014}. While the detailed behavior is somewhat uncertain and shows more complexity than a constant value of $\alpha$ that we consider here, Borg \& Morrissey do conclude that ``it would be difficult to throw a pitch that did not begin to rotate under the influence of aerodynamic shear'' \cite{Borg2014}.

In examining Equations~\ref{eq:psi} and \ref{eq:phi}, the size of the torque is dependent on two parameters, $\alpha$ and $v_0$. As we have stated, $\alpha$ is largely unconstrained but is likely of order $1/2$, while $v_0$ can be, and, indeed, is intentionally varied by the pitcher. Both of these parameters should increase the torque on the ball, leading to a larger nonlinear interaction between the ball's orientation and its spin. Effectively, increasing either $\alpha$ or $v_0$ is analogous to amplifying the nonlinearity in the problem that leads to the chaos observed in Type B models. To quantify the effect of both $\alpha$ and $v_0$ on the resulting chaotic dynamics, we selected a range of initial conditions with $\pi/5 \leq \psi \leq \pi/4$ rad and $0 \leq \omega_x \leq 6.0$ rad/s, while $\phi$ and $\omega_z$ are set to zero. We then set values for $\alpha$ and $v_0$ and run $10^5$ combinations of initial conditions. For each phase space sweep, we measure the largest Lyapunov exponent achieved in that range of initial conditions for specified values of $\alpha$ and $v_0$.

Figure~\ref{fig:lambdaDep} shows how the largest Lyapunov exponent achieved in the specified region of initial condition space varies with $\alpha$ and $v_0$. These numerical experiments demonstrate that the largest Lyapunov exponent is proportional to $\alpha^{0.527}$, tantalizingly close to $\sqrt{\alpha}$, and to $v_0^{0.882}$. This also demonstrates that for all non-zero values of the effective lever arm and all initial pitch speeds studied (and those used by knuckleball pitchers \cite{Nathan2012}) the behavior of our Type B model can be chaotic. 

Beyond simply identifying chaos, we can also quantify the timescale of the exponential divergence in phase space. For pitches to exhibit significant chaotic effects, the Lyapunov time scale $\tau_L = 1 / \lambda$ must be much shorter than what we term the flight time scale $\tau_F \approx y_0 / v_0$. When examining how the ratio $\tau_F / \tau_L$ varies with $\alpha$, we see that it varies from about 1 when $\alpha = 0.01$ to as much as 17.5 when $\alpha = 1$. If we set an arbitrary line requiring at least 5 e-foldings (a factor of $\approx 10^3$) to be possible during the ball's flight, then values of $\alpha$ as small as 0.15 attain this level of chaotic divergence. This then argues that only particularly tuned distributions of the asymmetry force can avoid substantial levels of chaotic behavior.

\begin{figure*}[t]
\includegraphics[width=0.49\linewidth]{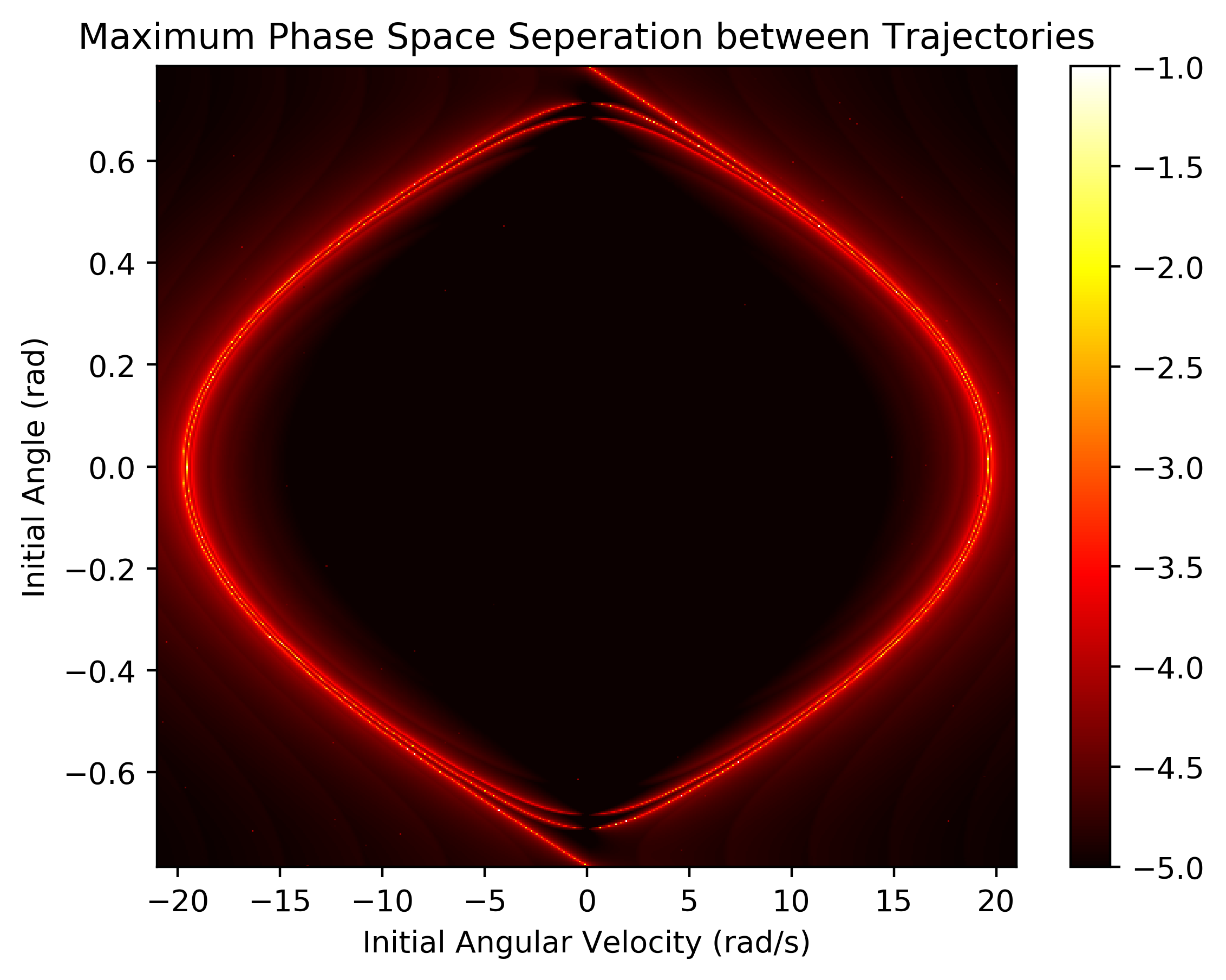}
\includegraphics[width=0.49\linewidth]{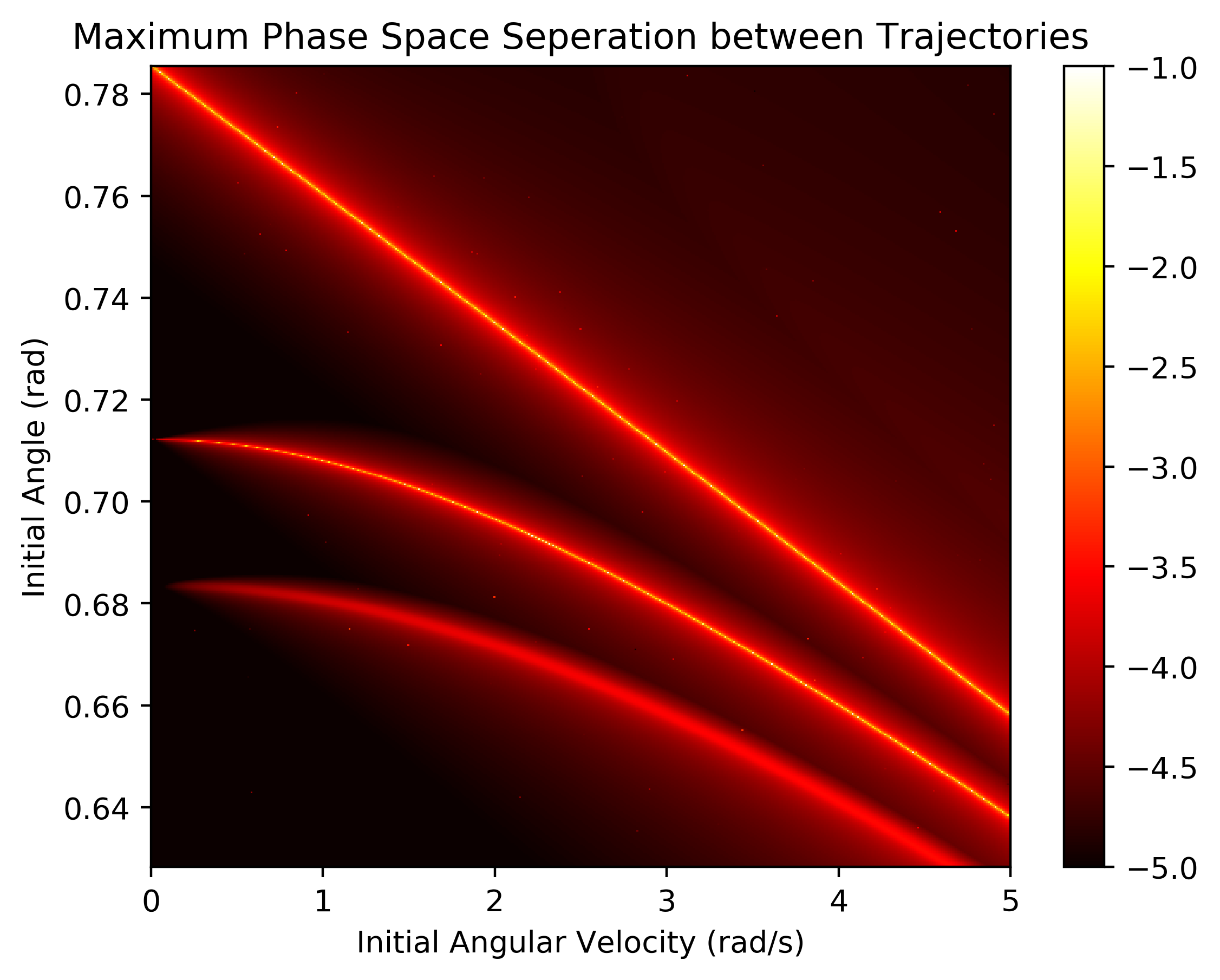}
\caption{\label{fig:PhaseSepMap} Left: Maximum phase space separation for Type B model pitches with $\phi = 0$, $\omega_z = 0$, $-\pi/4 \leq \psi \leq \pi/4$ rad, and $-21.0 \leq \omega_x \leq 21.0$ rad/s. Color indicates the base-10 logarithm of the maximum phase space separation for two initial conditions, one of which is initially perturbed by $10^{-6}$. Values range from black indicating non-chaotic behavior to red indicating marginal chaos to white indicating clearly chaotic dynamics. Chaotic behavior is achieved in a roughly circular band in parameter space. Right: Zoom-in of right-hand panel highlighting the region $\pi/5 \leq \psi \leq \pi/4$ rad and $0.0 \leq \omega_x \leq 5.0$ rad/s in order to show the fine structure of the band where chaotic behavior is achieved, which is confined to three distinct tracks through parameter space.}
\end{figure*}

If we now turn to the dependence of the largest Lyapunov exponent with $v_0$, we find an intriguing result. The nearly linear dependence of the largest Lyapunov exponent on $v_0$ leads to a roughly constant ratio of about 12 Lyapunov times per flight time, meaning that the initial phase space difference has time to grow by a factor of $e^{12}$ ($\approx 10^5$) on its journey to home plate independent of how fast the ball is thrown. This provides an interesting possible explanation for why knuckleballs seems to work quite well even though they are generally thrown much slower than other baseball pitches designed to take advantage of aerodynamic effects.

\subsection{Variation with Initial Orientation and Spin}

As we previously stated, when selecting initial angles and angular velocities at random we found that only 1.6\% produced exponentially growing separations in phase space. To explore range of possible behaviors as a function of the initial conditions, we now turn to a parameter sweep of possible initial angles and angular velocities while holding $v_0$ and $\alpha$ constant at 36 m/s and 0.5, respectively. In order to reduce the dimensionality of the space to explore, we chose to use the 2D form of the models. We explore the behavior of the system for $-\pi/4 \leq \psi \leq \pi/4$ rad, and $-21.0 \leq \omega_x \leq 21.0$ rad/s. Figure~\ref{fig:PhaseSepMap} shows the base-10 logarithm of the maximum phase space separation achieved as a function of initial angle and angular velocity for the entire region (left panel), as well as a zoom in on the region $\pi/5 \leq \psi \leq \pi/4$ rad, and $0.0 \leq \omega_x \leq 5.0$ rad/s (right panel). As expected from the symmetry of the problem, the results are anti-symmetric when $\psi \rightarrow -\psi$ and $\omega_x \rightarrow - \omega_x$. 

Chaotic behavior is confined to three distinct arcs in this cut through phase space. These arcs are clearly separate when studied at higher resolution (see right-hand panel). Along these ridges initial phase space separations on the order of $10^{-6}$ can grow five orders of magnitude or more in less than 0.5 s. All of the largest Lyapunov exponents shown in Figure~\ref{fig:lambdaDep} originate from initial conditions along one of these ridges. The logarithmic scaling of the color table in Figure~\ref{fig:PhaseSepMap} serves to highlight the narrowness of these ridges, with phase space separations dropping by orders of magnitude when moving even slightly away from the arcs.

The right-hand panel of Figure~\ref{fig:PhaseSepMap} shows a zoom-in on the region of the initial condition space that shows some of the largest divergence in possible trajectories. This is also the region where large variability is seen for initial rotation rates on the order of the ``less than a half turn of the ball on its way to the plate'' \cite{Clark2012}. Here changes in initial rotation rate of as little as a tenth of a radian per second -- imperceptible differences to even current high-speed photography -- can yield a change in the ball's position at the plate of as much as 10 cm. Similar behavior is seen in the horizontal motion of the ball when $\phi$ and $\omega_z$ are allowed to vary, meaning that the modeled behavior could yield several bat-widths of motion in both horizontal and vertical directions when imperceptible changes in the rotation rate and/or initial orientation of the ball are admitted.

\begin{figure}[t]
\centering
\includegraphics[width=0.6\linewidth]{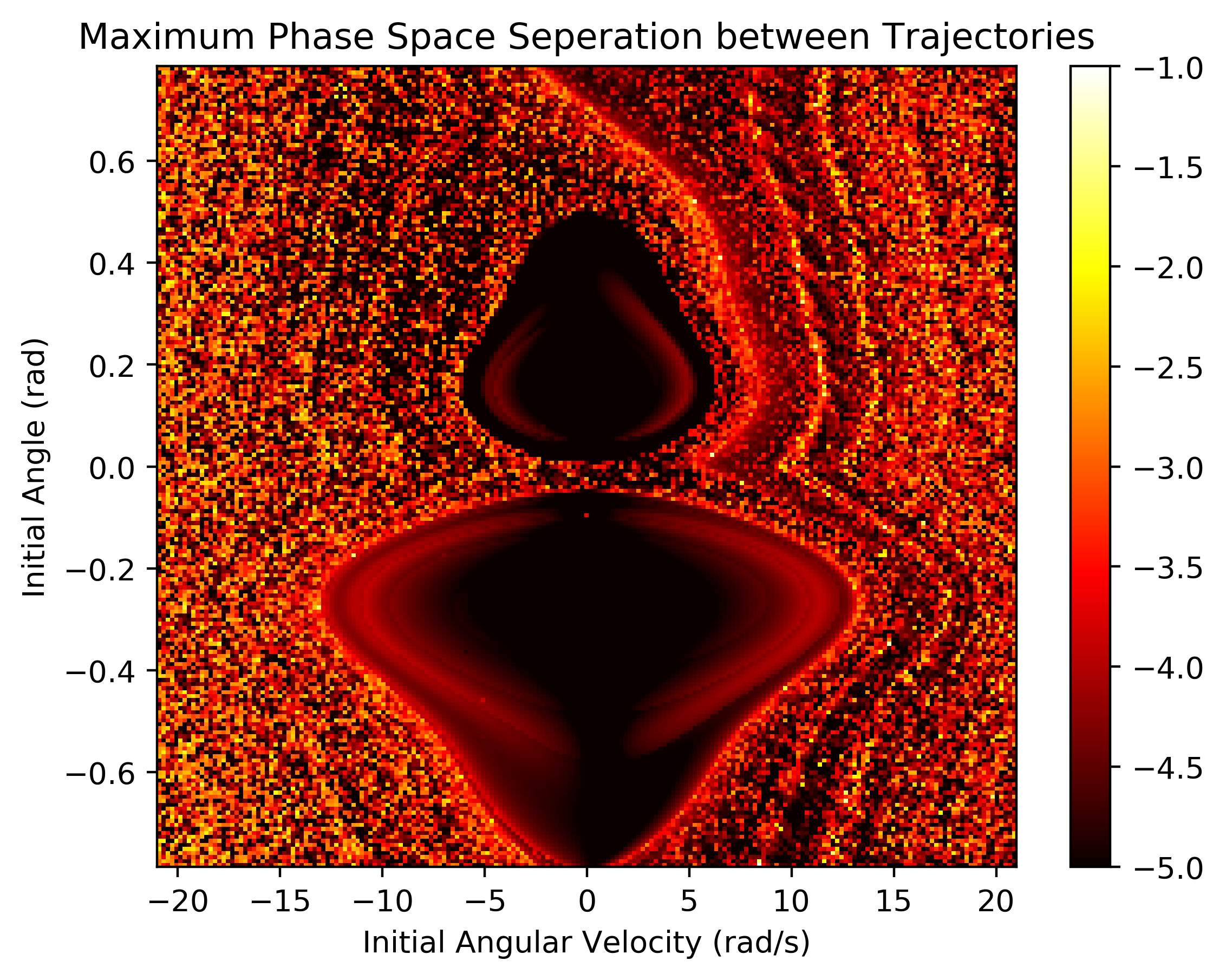}
\caption{\label{fig:PhaseSepMap_realCL} Similar to Fig.~\ref{fig:PhaseSepMap}, but for Type C models in the 4S configuration using empirical data for the angular dependence of the asymmetry force. Phase space separations $\Delta \chi$ range from a lower limits of approximately $10^{-3}$ (black) due to uncertainties in the empirical data, to as large as $\Delta \chi \gtrsim 0.3$ (white). Clearly Type C models display chaotic behavior over a much wider range of initial orientations and spins than Type B models. }
\end{figure}

Having found chaotic behavior in our Type B models, we might expect that the additional complexity of the Type C models would also yield chaotic dynamics. Indeed that does turn out to be the case. As in our Type B models, Type C models also show chaotic behavior for many choices of initial angle and angular velocity.  Figure~\ref{fig:PhaseSepMap_realCL} shows the maximum phase space separation $\Delta \chi$ for initial orientations and angular velocities, similar to those represented in Figure~\ref{fig:PhaseSepMap}. Roughly 12\% of models show $\Delta \chi > 10^{-3}$, which we interpret as clearly chaotic. In comparing these results with those seen for Type B models, we see some clear similarities in the presence and overall shape of chaotic tracks. We also see that Type C models have significant ranges in parameter space where no chaotic behavior is seen. As expected Type C models show even greater chaotic behavior than Type B models.

\section{Conclusion}

\begin{figure}[t]
\centering
\includegraphics[width=0.6\linewidth]{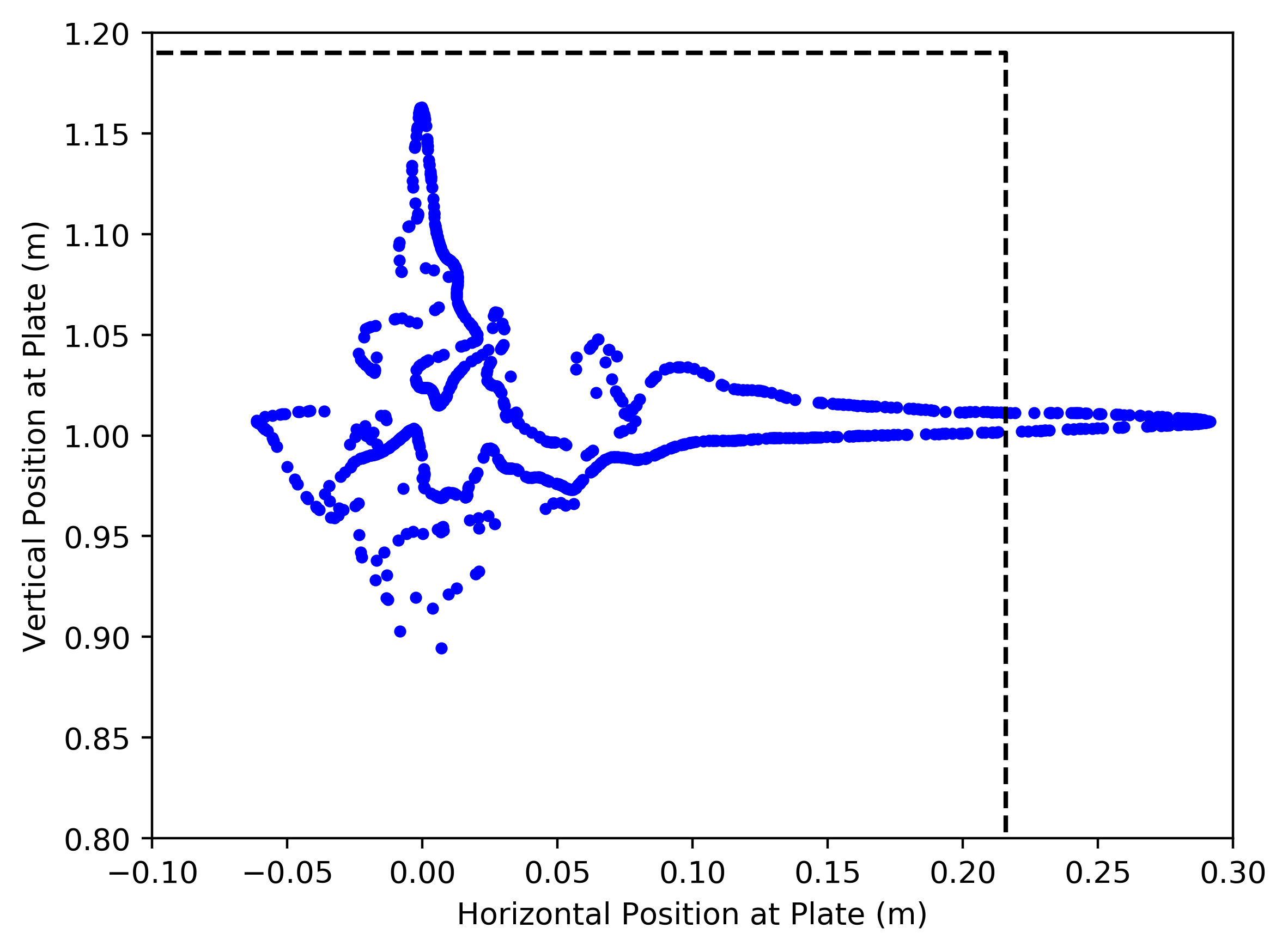}
\caption{\label{fig:PlateLocsHist} Scatter plot showing the locations at the plate for 1000 pitches using the Type C model. All pitches have initial orientations and angular velocities which vary by less than 0.01\%. For reference, the average initial conditions are given in Table~\ref{tab:ICs} with $\phi = -0.050$ rad, $\psi = 0.330$ rad, $\omega_x = 0.770$ rad/s, and $\omega_z = 1.948$ rad/s. Also shown for reference is the strike zone (black dotted line). The tracks through the plate location space show the range of plate locations achievable with practically identical pitches thanks to the chaotic nature of our Type C models. }
\end{figure}

It is perhaps unsurprising, given the professional baseball talent devoted to them, that knuckleballs should exhibit such complex and richly dynamic behavior even for the relatively simply models presented here. This is even more striking when compared to the reliability with which other pitches can be modeled \cite{Nathan2008}. In this paper we have presented simplified models that do not include dynamic coupling between the ball and the fluid flow, but rather treat aerodynamic forces simply as functions of the ball's velocity and orientation. We have shown that these models can mimic the strong variability seen in the location of a given pitch at the plate with very small changes in initial orientation and spin. We have also shown that our models that include torques on the ball due to the asymmetry force display dynamical chaos for certain regions of parameter space corresponding to the steepest gradients in plate location. Finally, we have also shown that the chaotic dynamics produced in these models is likely to produce significant uncertainty in the trajectory of the ball for realistic uncertainties in the initial conditions of the pitch. 

All three of our model types show the ability to produce variations in the location of the ball at the plate for small changes in the initial orientation and spin of the ball. Those variations are most pronounced for the Type C models where significant feedback on the angular velocity of the ball can lead to very dramatic changes in the position of the ball at the plate in both the vertical and horizontal directions. 

Our Type B and C models show strong sensitivity to initial conditions, leading to Lyapunov times as small as 26 ms. For Type B models, we have shown that the timescale for the exponential growth in uncertainties is strongly dependent on the effective lever arm of the torque from the asymmetry force, as well as on the velocity of the pitch. Interestingly, the near-linear dependence of the Lyapunov exponent on the initial velocity means that the magnitude of separation that can be achieved is nearly independent of the initial velocity. This may explain why the knuckleball is effective at lower speeds. The large Lyapunov exponents measured in our Type B and C models demonstrate that initial uncertainties in the orientation and spin of the ball can grow through as much as seven orders of magnitude over the ball's flight from the pitcher to home plate.

For pitchers and batters, this leads to the conclusion that predicting the trajectory of a knuckleball is not just difficult -- it is functionally impossible. Figure~\ref{fig:PlateLocsHist} shows the possible locations at which a knuckleball can arrive at home plate using our Type C model. All trajectories start with identical initial position and velocity, and with initial orientations and angular velocities that vary less that 0.01\%. This small variation in initial orientation and angular velocity grows as the pitches travel towards home plate, leading to the distribution shown. This produces a stark reminder that chaotic does not equate to random. The pitch considered cannot end up anywhere and indeed most of the available space is not accessible with these initial conditions. Chaotic does, however, mean that very small initial uncertainties can lead to very large uncertainties when the ball reaches the plate.

That physical knuckleballs exhibit some level of dynamical chaos has been speculated and is perhaps unsurprising \cite{Nathan2012}. What may be of greater interest is that large chaotic uncertainties can be realized with simplified models considered here. Adding additional nonlinearities and degrees of freedom in the models is not likely to reduce the chaotic behavior. Additionally, the level of uncertainty which can be achieved from chaotic behaviors shows that not only are knuckleballs chaotic, but that chaos may play an important role in making the pitch work. Exponential divergence on timescales as small as 26 ms and resulting movement of as much as 1.25 m means that not only are knuckleball models chaotic, but that the inherent chaotic uncertainty is much larger than the uncertainty in measurement achieved with modern tracking systems. In future work we hope to use the large numbers of knuckleballs tracked by Major League Baseball's StatCast system to confirm that real knuckleballs show chaotic dynamics similar to those seen in our models.

\section*{Acknowledgments}
We wish to acknowledge David Kagan for introducing us to the field and pointing us in the right direction, and John Borg for sharing his data. ES was funded by a Summer Research Award from the College of Natural Sciences at CSU, Chico.


\section*{References}

\bibliographystyle{elsarticle-num} 
\bibliography{knuckleball}

\begin{thebibliography}{10}
\expandafter\ifx\csname url\endcsname\relax
  \def\url#1{\texttt{#1}}\fi
\expandafter\ifx\csname urlprefix\endcsname\relax\def\urlprefix{URL }\fi
\expandafter\ifx\csname href\endcsname\relax
  \def\href#1#2{#2} \def\path#1{#1}\fi

\bibitem{Clark2012}
D.~Clark, {The Knucklebook: Everything You Need to Know about Baseball's
  Strangest Pitch}, 1st Edition, Ivan R. Dee, Chicago, USA, 2012.

\bibitem{Wakefield2011}
T.~Wakefield, T.~Massarotti, {Knuckler: My Life with Baseball's Most
  Confounding Pitch}, Houghton Mifflin Harcourt, Boston, 2011.

\bibitem{Dickey2012}
R.~A. Dickey, W.~Coffey, {Wherever I Wind Up}, Blue Rider Press, New York,
  2012.

\bibitem{Lorenz1963}
E.~N. Lorenz,
  \href{http://journals.ametsoc.org/jas/article-pdf/20/2/130/3406061/1520-0469}{{Deterministic
  Nonperiodic Flow}}, Journal of the Atmospheric Sciences 20~(2) (1963)
  130--141.
\newblock \href {https://doi.org/10.1175/1520-0469(1963)020<0130:dnf>2.0.co;2}
  {\path{doi:10.1175/1520-0469(1963)020<0130:dnf>2.0.co;2}}.
\newline\urlprefix\url{http://journals.ametsoc.org/jas/article-pdf/20/2/130/3406061/1520-0469}

\bibitem{Perc2005}
M.~Perc, \href{https://iopscience.iop.org/article/10.1088/0143-0807/26/4/003
  https://iopscience.iop.org/article/10.1088/0143-0807/26/4/003/meta}{{Visualizing
  the attraction of strange attractors}}, European Journal of Physics 26~(4)
  (2005) 579--587.
\newblock \href {https://doi.org/10.1088/0143-0807/26/4/003}
  {\path{doi:10.1088/0143-0807/26/4/003}}.
\newline\urlprefix\url{https://iopscience.iop.org/article/10.1088/0143-0807/26/4/003
  https://iopscience.iop.org/article/10.1088/0143-0807/26/4/003/meta}

\bibitem{Strogatz2015}
S.~H. Strogatz, {Nonlinear Dynamics and Chaos with Applications to Physics,
  Biology, Chemistry, and Engineering}, 2nd Edition, CRC Press, Boca Raton,
  Florida, 2015.

\bibitem{Brown2018}
R.~J. Brown, {A Modern Introduction to Dynamical Systems}, 1st Edition, Oxford
  University Press, Oxford, UK, 2018.

\bibitem{Nepomuceno2020}
E.~G. Nepomuceno, M.~Perc,
  \href{https://academic.oup.com/comnet/article/8/1/cnz015/5481207}{{Computational
  chaos in complex networks}}, Journal of Complex Networks 8~(1) (feb 2020).
\newblock \href {https://doi.org/10.1093/comnet/cnz015}
  {\path{doi:10.1093/comnet/cnz015}}.
\newline\urlprefix\url{https://academic.oup.com/comnet/article/8/1/cnz015/5481207}

\bibitem{Baseball2019}
M.~L. Baseball, {Baseball Savant} (2019).

\bibitem{Roegele2014}
J.~Roegele, {The Strike Zone During the PITCHf/x Era}, The Hardball Times
  Baseball Annual 2014 (2014) 1--12.

\bibitem{Watts1987}
R.~G. Watts, R.~Ferrer, {The lateral force on a spinning sphere: Aerodynamics
  of a curveball}, American Journal of Physics 55~(1) (1987) 40--44.
\newblock \href {https://doi.org/10.1119/1.14969} {\path{doi:10.1119/1.14969}}.

\bibitem{Nathan2008}
A.~M. Nathan, {The effect of spin on the flight of a baseball}, American
  Journal of Physics 76~(2) (2008) 119--124.
\newblock \href {https://doi.org/10.1007/978-0-387-46050-5_5}
  {\path{doi:10.1007/978-0-387-46050-5_5}}.

\bibitem{Nathan2012}
A.~M. Nathan, {Analysis of knuckleball trajectories}, Procedia Engineering 34
  (2012) 116--121.
\newblock \href {https://doi.org/10.1016/j.proeng.2012.04.021}
  {\path{doi:10.1016/j.proeng.2012.04.021}}.

\bibitem{Kagan2014}
D.~Kagan, A.~M. Nathan, {Simplified Models for the Drag Coefficient of a
  Pitched Baseball}, The Physics Teacher 52~(5) (2014) 278--280.
\newblock \href {https://doi.org/10.1119/1.4872406}
  {\path{doi:10.1119/1.4872406}}.

\bibitem{EscaleraSantos2019}
G.~J. {Escalera Santos}, M.~A. Aguirre-L{\'{o}}pez,
  O.~D{\'{i}}az-Hern{\'{a}}ndez, F.~Hueyotl-Zahuantitla, J.~Morales-Castillo,
  F.-J. Almaguer, {On the Aerodynamic Forces on a Baseball, With Applications},
  Frontiers in Applied Mathematics and Statistics 4~(January) (2019) 1--9.
\newblock \href {https://doi.org/10.3389/fams.2018.00066}
  {\path{doi:10.3389/fams.2018.00066}}.

\bibitem{Adair2002}
R.~K. Adair, {The Physics of Baseball}, 3rd Edition, Harper Perennial, New
  York, 2002.

\bibitem{Watts1975}
R.~Watts, E.~Sawyer, {Aerodynamics of a knuckleball}, American Journal of
  Physics 43~(11) (1975) 960--964.

\bibitem{Morrissey2009}
M.~P. Morrissey,
  \href{http://epublications.marquette.edu/theses{\_}open/8}{{The aerodynamics
  of the knuckleball pitch: An experimental investigation into the effects that
  the seam and slow rotation have on a baseball}}, Ph.D. thesis, Marquette
  University (2009).
\newline\urlprefix\url{http://epublications.marquette.edu/theses{\_}open/8}

\bibitem{Asai2011}
T.~Asai, K.~Kamemoto,
  \href{http://dx.doi.org/10.1016/j.jfluidstructs.2011.03.016}{{Flow structure
  of knuckling effect in footballs}}, Journal of Fluids and Structures 27~(5-6)
  (2011) 727--733.
\newblock \href {https://doi.org/10.1016/j.jfluidstructs.2011.03.016}
  {\path{doi:10.1016/j.jfluidstructs.2011.03.016}}.
\newline\urlprefix\url{http://dx.doi.org/10.1016/j.jfluidstructs.2011.03.016}

\bibitem{Higuchi2012}
H.~Higuchi, T.~Kiura,
  \href{http://dx.doi.org/10.1016/j.jfluidstructs.2012.01.004}{{Aerodynamics of
  knuckle ball: Flow-structure interaction problem on a pitched baseball
  without spin}}, Journal of Fluids and Structures 32 (2012) 65--77.
\newblock \href {https://doi.org/10.1016/j.jfluidstructs.2012.01.004}
  {\path{doi:10.1016/j.jfluidstructs.2012.01.004}}.
\newline\urlprefix\url{http://dx.doi.org/10.1016/j.jfluidstructs.2012.01.004}

\bibitem{Borg2014}
J.~P. Borg, M.~P. Morrissey, {Aerodynamics of the knuckleball pitch:
  Experimental measurements on slowly rotating baseballs}, American Journal of
  Physics 82~(10) (2014) 921--927.
\newblock \href {https://doi.org/10.1119/1.4885341}
  {\path{doi:10.1119/1.4885341}}.

\bibitem{Texier2016}
B.~D. Texier, C.~Cohen, D.~Qu{\'{e}}r{\'{e}}, C.~Clanet, {Physics of
  knuckleballs}, New Journal of Physics 18~(7) (2016).
\newblock \href {https://doi.org/10.1088/1367-2630/18/7/073027}
  {\path{doi:10.1088/1367-2630/18/7/073027}}.

\bibitem{Aguirre-Lopez2017}
M.~A. Aguirre-L{\'{o}}pez, O.~D{\'{i}}az-Hern{\'{a}}ndez, F.~J. Almaguer,
  J.~Morales-Castillo, G.~J. {Escalera Santos}, {A phenomenological model for
  the aerodynamics of the knuckleball}, Applied Mathematics and Computation 311
  (2017) 58--65.
\newblock \href {https://doi.org/10.1016/j.amc.2017.05.001}
  {\path{doi:10.1016/j.amc.2017.05.001}}.

\bibitem{Community2019}
T.~S. Community, {SciPy Reference Guide v1.3.0} (2019).

\bibitem{Dormand1980}
J.~R. Dormand, P.~J. Prince, {A family of embedded Runge-Kutta formulae},
  Journal of Computational and Applied Mathematics 6~(1) (1980) 19--26.
\newblock \href {https://doi.org/10.1016/0771-050X(80)90013-3}
  {\path{doi:10.1016/0771-050X(80)90013-3}}.

\end{thebibliography}

\end{document}